\documentclass[12pt]{article}
\usepackage{amsmath}
\usepackage{graphicx,psfrag,epsf,color,subfigure}
\usepackage{enumerate}
\usepackage{natbib}

\usepackage[unicode=true,pdfusetitle,
   bookmarks=true, 
   bookmarksnumbered=false, 
   bookmarksopen=false,
   breaklinks=false, 
   pdfborder={0 0 1}, 
   backref=false, 
   colorlinks=false]{hyperref}

\usepackage{amssymb}
\usepackage{multirow}
\usepackage{array}
\usepackage{bm}
\usepackage{booktabs}
\usepackage{float}
\usepackage[font=small,labelfont=bf]{caption}
\setcitestyle{citesep={;}}
\usepackage[total={6.4in,8.6in}]{geometry}
\usepackage{algorithm2e} 

\newcommand{\Mean}{{\mathbb{E}}}
\newcommand{\Var}{{\mbox{Var}}}
\newcommand{\Cov}{{\mbox{cov}}}

\newcommand{\prob}{{\mbox{Pr}}}
\newcommand\independent{\protect\mathpalette{\protect\independenT}{\perp}}
\def\independenT#1#2{\mathrel{\rlap{$#1#2$}\mkern2mu{#1#2}}}
\let\code=\texttt
\let\proglang=\textsf 

\usepackage{authblk}

\newtheorem{theorem}{Theorem}[section]

\newtheorem{lemma}{Lemma}[section]

\DeclareMathOperator*{\argmax}{arg\,max}

\begin{document}

\def\spacingset#1{\renewcommand{\baselinestretch}%
{#1}\small\normalsize} \spacingset{1}
%
%
\title{\bf Calibrated Optimal Decision Making with Multiple Data Sources and Limited Outcome}
\author[1]{Hengrui Cai   \thanks{hengrc1@uci.edu} } 
\author[2]{Wenbin Lu\thanks{wlu4@ncsu.edu}} 
\author[2]{Rui Song\thanks{rsong@ncsu.edu}}
\affil[2]{Department of Statistics, University of California Irvine} 
\affil[2]{Department of Statistics, North Carolina State University} 
 \date{}
 \maketitle  

\baselineskip=21pt

%


  
\begin{abstract}

We consider the optimal decision-making problem in a primary sample of interest with multiple auxiliary sources available. The outcome of interest is limited in the sense that it is only observed in the primary sample. In reality, such multiple data sources may belong to {heterogeneous studies} 
and thus cannot be combined directly. This paper proposes {a new framework to handle heterogeneous samples and address the limited outcome simultaneously through a novel {c}alibrated  {o}ptimal {d}ecision-m{a}king method}, by leveraging the common intermediate outcomes in multiple data sources. 
{Specifically, our method allows the baseline covariates across different samples to have either homogeneous or heterogeneous distributions.}  
Under the equal conditional means of intermediate outcomes in different samples given baseline covariates and the treatment information, 
we show that the proposed estimator of the conditional mean outcome is asymptotically normal and more efficient than using the primary sample solely. 
Extensive experiments on simulated datasets demonstrate empirical validity and improved efficiency using our approach, followed by a real application to electronic health records.

\end{abstract}

   \section{Introduction} \label{sec:1}
Personalized decision-making is an emerging artificial intelligence paradigm tailored to an individual's characteristics, with wide real-world applications in 
precision medicine \citep{chakraborty2013statistical}. %
 The ultimate goal is to optimize the outcome of interest by assigning the right treatment to the right subjects. The resulting best strategy is referred to as the optimal decision rule, ODR.
A large number of approaches have been developed for finding ODR, including Q-learning \citep{watkins1992q,chak2010,qian2011performance}, A-learning \citep{murphy2003optimal,robins2004optimal}, value search methods \citep{zhang2012robust,wang2018quantile,nie2020learning}, and decision tree-based methods \citep{laber2015tree,zhangyc2016}. 
However, all these methods are developed based on data from a single source where the primary outcome of interest can be observed for all subjects, making these works less practical in more complicated situations.

There are many applications involving multiple datasets from different sources, where the primary outcome of interest is limited in the sense that it is only observed in some data sources. 
Take the treatment of sepsis as an instance. In the Medical Information Mart for Intensive Care (MIMIC-III) clinical database \citep{goldberger2000physiobank,johnson2016mimic,biseda2020prediction}, thousands of patients in intensive care units of the Beth Israel Deaconess Medical Center between 2001 and 2012 were treated with different medical supervisions such as the {vasopressor} and followed up for the mortality due to sepsis as the primary outcome of interest. 
%
In addition, we can observe other post-treatment intermediate outcomes (also known as surrogacies or proximal outcomes), such as the total urine output and the cumulated net of metabolism. These intermediate outcomes, as well as baseline variables and the treatment information collected in the MIMIC-III data, were also recorded in the electronic Intensive Care Units (eICU) collaborative research database \citep{goldberger2000physiobank,pollard2018eicu} that contains over 200,000 admissions to intensive care units across the United States between 2014 and 2015. 
Yet, the outcome of interest was not reported in eICU. Hence, we view the MIMIC-III data as the primary sample and the eICU data as the auxiliary sample without the outcome of interest, 
leading to one desideratum in precision medicine on finding ODR to optimize the mortality rate of sepsis based on these datasets. 

However, integrating multiple data sources  {from heterogeneous studies} for deriving ODR can be particularly challenging.  For instance, the MIMIC-III and eICU data were collected from {different locations during different periods}. These two samples show certain heterogeneity (see details in Section \ref{sec:5}) such as diverse probability distributions in baseline covariates, the treatment, and intermediate outcomes, and thus cannot be combined directly. 
In this work, we propose \textcolor{black}{a new framework to handle heterogeneous samples and address the limited outcome simultaneously via a novel {c}alibrated {o}ptimal {d}ecision m{a}king method, namely CODA.} 
Motivated by the common data structures and similar data dependences in the MIMIC-III and eICU data,  
CODA naturally utilizes the common intermediate outcomes in multiple data sources via a calibration technique \citep{chen2000unified,chen2002cox,cao2009improving,lumley2011connections}, to improve the performance of the treatment decision rule by borrowing information from auxiliary samples.  
 
Our contributions can be summarized as follows. 
First, 
to safely borrow information from auxiliary samples, we {propose the comparable intermediate outcomes assumption}, that is, the conditional means of intermediate outcomes given baseline covariates and the treatment information are the same in the two samples. This assumption avoids the specification of the missing mechanism in auxiliary samples and is more practically suitable for heterogeneous studies. {Second, all the current calibration-based methods require that covariates in the primary and auxiliary samples are from the same distribution. In this work, we allow baseline covariates across different studies to have either homogeneous or heterogeneous distributions. We propose a unified framework for deriving a calibrated doubly robust estimator of the conditional mean outcome of interest (known as the value function) through the projection onto the difference of value estimators for common intermediate outcomes in the two samples. 
When the distributions of baseline covariates differ in different samples, we construct the calibrated value estimator by rebalancing the value estimators of common intermediate outcomes in two samples based on their posterior sampling probability. Third, to handle the large-scale datasets (such as the MIMIC-III and eICU data) and obtain interpretable decision rules, we develop an iterative policy tree search algorithm to find the decision rule that maximizes the calibrated value estimator in the primary sample.} 
 {Fourth, we establish the asymptotic normality of the calibrated value estimator under the estimated ODR obtained by CODA for both homogeneous and heterogeneous covariates,  
which is shown to be more efficient than that obtained using the primary sample solely. }

\subsection{Related Works}\label{sec:related}

There are several recent works in using multiple data sources for estimating the average treatment effect \citep{yang2019combining,athey2020combining, kallus2020role} or deriving a robust ODR to account for heterogeneity in multiple data sources \citep{shi2018maximin,mo2020learning}. However, the settings and goals of these studies are different from what we consider here. 
{Specifically, in the works of \cite{yang2019combining}, \cite{athey2020combining}, and  \cite{kallus2020role}, it was assumed that the two samples are from the same population and were linked together through a missing data framework, such as under the missing-at-random assumption. This allows to either develop a calibrated estimator using the common baseline covariates in both samples \citep{yang2019combining} or} impute the missing primary outcome in the auxiliary data \citep{athey2020combining, kallus2020role}, so that a more efficient estimator can be constructed for the average treatment effect. Whereas, 
the multiple data sources considered in our study may come from {heterogeneous studies} as in the MIMIC III and eICU data, and hence their missing data framework cannot be directly applied in our problem. For example, simply extending the calibration method for the average treatment effect considered in  \cite{yang2019combining} may lead to a biased result when the covariate distributions in two samples are heterogeneous; while the adaption of the methods of \cite{athey2020combining} and  \cite{kallus2020role} requires an untestable assumption that the conditional means of the outcome  given baseline covariates, treatment, and intermediate outcomes are the same across samples. 
On the other hand, the main goal of \cite{shi2018maximin} and \cite{mo2020learning} is to develop a single ODR that can work for multiple data sources with heterogeneity in data distributions or outcome models, and their methods do not allow missingness in outcomes. In contrast, we are interested in safely improving the efficiency of value of ODR for the limited outcome, 
by leveraging available auxiliary data sources.

\subsection{Outline of the Paper}
The rest of this paper is organized as follows. We introduce notations and assumptions in Section \ref{sec:2}. In Section \ref{sec:3}, we propose {two calibrated optimal decision-making methods, CODA-HO and CODA-HE, for homogeneous and heterogeneous covariates, respectively, and detail their implementation based on the iterative policy tree search algorithm}. All the theoretical properties are established in Section \ref{sec:thms}. 
Extensive simulations are conducted to demonstrate the empirical performance of the proposed method in Section \ref{sec:4}, followed by a real application in developing ODR for treating sepsis using the MIMIC-III data as the primary sample and the eICU  data as the auxiliary sample in Section \ref{sec:5}. We conclude our paper in Section \ref{sec:6}. The technical proofs are given in the Appendix. The source code is publicly available at our repository at \url{https://github.com/HengruiCai/CODA} implemented in \textsf{R} language.

 \section{Statistical Framework} \label{sec:2}
For simplicity of exposition, we consider a study with two data sources.
Suppose there is a primary sample of interest $P$. Let $X_P=[X_P^{(1)},\cdots, X_P^{(r)}]^\top$ denote $r$-dimensional individual's baseline covariates with the support $\mathbb{X}_P \subseteq \mathbb{R}^r$, and $A_P\in \{0,1\}$ denote the binary treatment an individual receives. After a treatment $A_P$ is assigned, we first obtain $s$-dimensional intermediate outcomes $M_P=[M_P^{(1)},\cdots, M_P^{(s)}]^\top$ with support $\mathbb{M}_P \subseteq \mathbb{R}^s$, and then observe the primary outcome of interest $Y_P $ with support $\mathbb{Y}_P \subseteq \mathbb{R}$, the larger the better by convention. Denote $N_P$ as the sample size for the primary sample, which consists of $\{P_i=(X_{P,i},A_{P,i},M_ {P,i},Y_ {P,i}), i = 1, \dots , N_P\}$ independent and identically distributed across $i$. 
To gain efficiency, we include an auxiliary sample $U$ available from another source. The auxiliary sample $U$ contains {the same set of baseline covariates $X_U=[X_U^{(1)},\cdots, X_U^{(r)}]^\top$ (with the same ordering as $X_P$ when $r>1$), the treatment $A_U$, and intermediate outcomes $M_U=[M_U^{(1)},\cdots, M_U^{(s)}]^\top$ (with the same ordering as $M_P$ when $s>1$) as in the primary sample}, with the support $\mathbb{X}_U \subseteq \mathbb{R}^r$, $\{0,1\}$, and $\mathbb{M}_U \subseteq \mathbb{R}^s$, respectively. Yet, the outcome of interest is limited in the primary sample and is not available in the auxiliary sample. Let $N_U$ denote the sample size for the independent and identically distributed auxiliary sample that includes $\{U_i=(X_{U,i},A_{U,i},M_ {U,i}), i = 1, \dots , N_U\}$. Denote $t={N_P/ N_U}$ as the sample ratio between the primary and the auxiliary sample, and $0<t<+\infty$. {We allow the distributions of baseline covariates, treatments, and intermediate outcomes differ in two samples.}
  
In the primary sample, define the potential outcomes $Y_P^*(0)$ and $Y_P^*(1)$ as the outcome of interest that would be observed after an individual receiving treatment 0 or 1, respectively. Similarly, we define the potential outcomes $\{M_P^*(0),M_P^*(1)\}$ and $\{M_U^*(0),M_U^*(1)\}$ as intermediate outcomes that would be observed after an individual receiving treatment 0 and 1 for the primary sample and the auxiliary sample, respectively. Define the propensity score function as the conditional probability of receiving treatment 1 given baseline covariates as $x$, denoted as $\pi_P(x)=\prob(A_P=1\mid X_P=x)$ for the primary sample and $\pi_U(x)=\prob(A_U=1\mid X_U=x)$ for the auxiliary sample. 
A decision rule is a deterministic function $d(\cdot)$ that maps covariate space $\mathbb{X}_P$ to the treatment space $\{0,1\}$. Define the potential outcome of interest under $d(\cdot)$ as 
	$Y_P^*(d)=Y_P^*(0) \{1-d(X_P)\}+Y_P^*(1) d(X_P),$ 
which would be observed if a randomly chosen individual from the primary sample had received a treatment according to $d(\cdot)$, where we suppress the dependence of $Y_P^*(d)$ on $X_P$. The value function under $d(\cdot)$ is defined as the expectation of the potential outcome of interest over the primary sample as
$V(d)=\Mean \{Y_P^*(d)\}=\Mean [Y_P^*(0) \{1-d(X_P)\}+Y_P^*(1) d(X_P)].$ 
As a result, the optimal decision rule (ODR)  for the primary outcome of interest is defined as the maximizer of the value function among a class of decision rules $\Pi$,
$d^{opt} =\argmax_{d \in \Pi} V(d).$ 
Similarly, we define the potential intermediate outcomes under $d(\cdot)$ for two samples as  
	$M_P^*(d)=M_P^*(0) \{1-d(X_P)\}+M_P^*(1) d(X_P)$ and 
	  $M_U^*(d)=M_U^*(0) \{1-d(X_U)\}+M_U^*(1) d(X_U).$ 
Here, $M_P^*(d)$ and $M_U^*(d)$ are $s\times 1$ vectors if $s>1$. 
To identify ODR for the primary outcome of interest from observed data, as standard in the causal inference literature \citep{rubin1978bayesian}, we make the following assumptions:




\noindent \textbf{(A1)}. Stable Unit Treatment Value Assumption: $M_P= A_P M_P^*(1)  + (1-A_P)M_P^*(0);$\\ 
$Y_P= A_P Y_P^*(1)  + (1-A_P)Y_P^*(0); \quad
M_U= A_U M_U^*(1)  + (1-A_U)M_U^*(0).$\\ 
\noindent \textbf{(A2)}. Ignorability: $\{M_P^*(0),M_P^*(1), Y_P^*(0),Y_P^*(1)\} \independent A_P\mid X_P; \\   \{M_U^*(0),M_U^*(1)\}\independent A_U\mid X_U.$ \\
 \noindent \textbf{(A3)}. Positivity: There exist constants $c_{1}$, $c_{2}$, $c_{3}$, and $c_{4}$ such that with probability 1, $0<c_{1}\leq\pi_P(x)\leq c_{2}<1$ for all $x \in \mathbb{X}_P$, and $0<c_{3}\leq\pi_U(x)\leq c_{4}<1$ for all $x \in \mathbb{X}_U$.
 
The above (A1) to (A3) are standard in personalized decision-making \citep[see][]{zhang2012robust,wang2018quantile,nie2020learning}, to guarantee that the value function of intermediate outcomes in two samples and the value of the outcome of interest in the primary sample are estimable from observed data. See more discussions of (A2) when the sets of $X_P$ and $X_U$ are different in Section \ref{sec:6}. 
Next, we make an assumption on the conditional means of intermediate outcomes to connect different data sources as follows.

\noindent \textbf{(A4)}. Comparable Intermediate Outcomes Assumption: \\
$\Mean(M_P\mid X_P=x, A_P=a)=\Mean(M_U\mid X_U=x,A_U=a), \text{ for all } x\in{\mathbb{X}_P\cup\mathbb{X}_U} \text{ and } a\in\{0,1\}.$

The above assumption states that the conditional means of intermediate outcomes given baseline covariates $x$ and the treatment information $a$ are the same in the two samples, {for all $x$ and $a$ in the union of the supports of the two samples}. 
This assumption automatically holds when the data sources are from the same population, where $\{X_P,A_P,M_P\}$ has the same probability distribution of $\{X_U,A_U,M_U\}$, as commonly assumed in the literature \citep[see][and more details in Section \ref{sec:related}]{yang2019combining,athey2020combining, kallus2020role}.  It is also worthy to mention that (A4) is  testable based on observed data. 
For example, one can test the equality of two conditional mean models based on some posited parametric regression models 
such as linear regression \citep{chow1960tests} or non-linear regression \citep{mahmoudi2018testing}. 
In addition, we consider the class $\Pi$ that satisfies the following condition.

\noindent \textbf{(A5).} $\Pi$ has a finite Vapnik-Chervonenkis dimension and is countable.

Assumption (A5) is commonly used in statistical learning and empirical process theory \citep[see][]{kitagawa2018should,rai2018statistical}. 
Popular classes of decision rules that satisfy (A5) include finite-depth decision trees  \citep{athey2017efficient} and parametric decision rules \citep{zhang2012robust}. 
   {To handle the large-scale datasets motivated by MIMIC-III and eICU data and obtain interpretable decision rules, we consider a class of decision trees with finite-depth $L_n \leq \kappa \log_2(n)$ for some $\kappa < 1/2$, denoted as $\Pi_1$, and search ODR within $\Pi_1$.}

\section{Method}\label{sec:3}
%
%
%

\subsection{Calibrated Optimal Decision Making for Homogeneous Baseline Covariates}\label{sec:CODA_homo}
{We first focus on the case where the distributions of baseline covariates in different samples are the same, $X_P \sim X_U$. Consider} the doubly robust (DR) estimators \citep{zhang2012robust} for the outcome of interest in the primary sample and intermediate outcomes in the two samples. Specifically, the DR estimator for the outcome of interest in the primary sample is   
 \begin{equation*} 
 \begin{split}
\widehat{V}_{P}(d)=   {1\over N_P}\sum_{i=1}^{N_P}  {\mathbb{I}\{A_{P,i}=d(X_{P,i})\}  [Y_{P,i} - \widehat{\mu}_P\{X_{P,i},d(X_{P,i})\} ]\over{A_{P,i} \widehat{\pi}_P (X_{P,i})+(1-A_{P,i})\{1-\widehat{\pi}_P(X_{P,i})\}}} + \widehat{\mu}_P\{X_{P,i},d(X_{P,i})\} ,
\end{split}
\end{equation*}
where $\widehat{\pi}_P$ is the estimator of the propensity score function, and $\widehat{\mu}_P(x,a)$ is the estimated conditional mean for ${\mu}_P(x,a)\equiv \Mean (Y_P\mid X_P=x,A_P=a)$, in the primary sample. Following arguments in \cite{zhang2012robust,luedtke2016statistical,kitagawa2018should,rai2018statistical}, 
we have the asymptotic normality for the value estimator as
\begin{equation}\label{lem1}
\sqrt{N_P} \Big\{\widehat{V}_{P}(d)  -  V(d) \Big\}  \rightsquigarrow  N\Big\{0, \sigma_Y^2(d)\Big\},
\end{equation}
where $\sigma_Y^2(d)$ is the asymptotic variance given any  $d(\cdot)$.
{Next, 
we introduce the calibrated value estimator. 
By assumption (A4) and 
$X_P\sim X_U $, we establish
the following lemma.}

 {\begin{lemma}\label{lem_cio}
Under assumptions (A1) - (A4) and $X_P\sim X_U $, we have
\begin{equation*}
\begin{split}
\Mean \{M_P^*(d)\}&=\Mean [M_P^*(0) \{1-d(X_P)\}+M_P^*(1) d(X_P)] \\
&= \Mean [M_U^*(0) \{1-d(X_U)\}+M_U^*(1) d(X_U)] = \Mean \{M_U^*(d)\}.
\end{split}
\end{equation*}
\end{lemma}
The detailed proof of Lemma \ref{lem_cio} is provided in the Appendix. Based on Lemma \ref{lem_cio}, the value functions for the intermediate potential outcomes under $d(\cdot)$ in the two samples are the same, that is,
$W_P(d)=\Mean \{M_P^*(d)\}=\Mean \{M_U^*(d)\}= W_U(d)\equiv W(d),$ 
where $W_P(d)$, $W_U(d)$, $W(d)$ are $s\times 1$ value vectors when $s>1$. This motivates us to derive the calibrated value estimator by projecting the value estimator of the outcome of interest in the primary sample on the differences of the value estimators of intermediate outcomes in the two samples.} 
%
Following assumption (A4), we define the conditional mean of intermediate outcomes as $\theta(x,a) \equiv \Mean(M_P\mid X_P=x, A_P=a)=\Mean(M_U\mid X_U=x,A_U=a)$, which is a $s\times 1$ vector. Then, we the DR value estimators for intermediate outcomes in two samples is
\begin{equation*} 
 \begin{split}
&\widehat{W}_{P}(d)=   {1\over N_P}\sum_{i=1}^{N_P} {\mathbb{I}\{A_{P,i}=d(X_{P,i})\}[M_{P,i}  - \widehat{\theta} \{X_{P,i},d(X_{P,i})\} ]\over{A_{P,i} \widehat{\pi}_P (X_{P,i})+(1-A_{P,i})\{1-\widehat{\pi}_P(X_{P,i})\}}}  + \widehat{\theta} \{X_{P,i},d(X_{P,i})\} ,\\
&\widehat{W}_{U}(d)=   {1\over N_U}\sum_{i=1}^{N_U} {\mathbb{I}\{A_{U,i}=d(X_{U,i})\} [M_{U,i}  - \widehat{\theta} \{X_{U,i},d(X_{U,i})\} ]\over{A_{U,i} \widehat{\pi}_U (X_{U,i})+(1-A_{U,i})\{1-\widehat{\pi}_U(X_{U,i})\}}} + \widehat{\theta}\{X_{U,i},d(X_{U,i})\}  ,
\end{split}
\end{equation*}
where $\widehat{W}_{P}(d)$ and $\widehat{W}_{U}(d)$ are $s\times 1$ vectors, $\widehat{\pi}_U $ is the estimator of the propensity score in the auxiliary sample, and $\widehat{\theta}(x,a)$ is the estimated conditional mean  function for $\theta(x,a)$ based on two samples combined under assumption (A4). Similarly, 
we have 
\begin{equation*}\label{lem2}
\sqrt{N_P} \Big\{\widehat{W}_{P}(d)  -  W(d) \Big\} \rightsquigarrow N_s\Big\{\boldsymbol{0}_{s}, \Sigma_P(d)\Big\},\quad \sqrt{N_U} \Big\{\widehat{W}_{U}(d)  -  W(d) \Big\} \rightsquigarrow N_s\Big\{\boldsymbol{0}_{s}, \Sigma_U(d)\Big\},
\end{equation*}
where $\boldsymbol{0}_{s}$ is the $s$-dimensional zero vector, $\Sigma_P$ and $\Sigma_U$ are $s\times s$ matrices presenting the asymptotic covariance matrices for two samples, and $ N_s(\cdot,\cdot)$ is the $s$-dimensional multivariate normal distribution. {According to Lemma \ref{lem_cio}, both $\widehat{W}_{P}(d)$ and $\widehat{W}_{U}(d)$ converge to the same value function $W(d) $.}
The following lemma establishes the asymptotic distribution of the differences of the value estimators of
intermediate outcomes in the two samples, under some technical conditions (A6) and (A7) detailed in Section \ref{sec:thms}. 
\textcolor{black}{ \begin{lemma}\label{lem3}
Under conditions (A1)-(A6), (A7. i, ii, and iii), and $X_P\sim X_U $, we
have
\begin{equation*}
\sqrt{N_P} \Big\{\widehat{W}_{P}(d)  - \widehat{W}_{U}(d) \Big\}  \rightsquigarrow  N_s\Big\{\boldsymbol{0}_{s}, \Sigma_M(d)\Big\},
\end{equation*}
where $\Sigma_M(d) = \Sigma_P(d)+ T\Sigma_U(d)$ is a $s\times s$ covariance matrix, and $T\equiv \lim_{N_P\to +\infty} t \in(0, +\infty)$.
\end{lemma}}
 The key gradient to prove Lemma \ref{lem3} lies in the fact that the two samples ($P$ and $U$) are collected from two different independent sources. The proof of Lemma \ref{lem3} is given in the Appendix. Based on Lemma \ref{lem3}, the asymptotic covariance matrix $\Sigma_M(d)$ is a weighted sum of the asymptotic covariance from each sample, where the weight is determined by the limiting sample ratio between two samples. 
%
%
%
Based on the results established in \eqref{lem1} and Lemma \ref{lem3}, we have the following asymptotic joint distribution  
 \begin{equation*}
  \begin{split}
&\sqrt{N_P}		 \begin{bmatrix}
   			\widehat{V}_{P}(d)  -  V(d) \\
   			\widehat{W}_{P}(d)  - \widehat{W}_{U}(d) 
			\end{bmatrix}
	 \rightsquigarrow N_{s+1} \Bigg\{\boldsymbol{0}_{s+1}, \begin{bmatrix}
   			\sigma_Y^2(d) , \boldsymbol{\rho}(d)^\top\\
   			\boldsymbol{\rho}(d), \Sigma_M(d)
			\end{bmatrix}\Bigg\},    \text{ for all } d(\cdot),
\end{split}
\end{equation*}
where $\boldsymbol{\rho}(d)$ is the $s\times 1$ asymptotic covariance vector between the value estimator of the outcome of interest in the primary sample and the differences of the value estimators of intermediate outcomes between two samples. 
It follows that the conditional distribution of $\sqrt{N_P}\{\widehat{V}_{P}(d)  -  V(d)\}$ given the estimated value differences of intermediate outcomes as
 \begin{equation}\label{eqn:proj}
  \begin{split}
 &\sqrt{N_P}\Big[\widehat{V}_{P}(d)  -  V(d) - \boldsymbol{\rho}(d)^\top  \Sigma_M^{-1}(d)  \{\widehat{W}_{P}(d)  - \widehat{W}_{U}(d)  \}\Big] \mid  \sqrt{N_P}\Big\{\widehat{W}_{P}(d)  - \widehat{W}_{U}(d) \Big\} \\
& \rightsquigarrow N\Big\{0, \sigma_Y^2(d)  - \boldsymbol{\rho}(d)^\top \Sigma_M^{-1} (d) \boldsymbol{\rho}(d)  \Big\} , \quad \text{ for all } d(\cdot).
\end{split}
\end{equation}
From \eqref{eqn:proj}, by projecting the value estimator of the outcome of interest on the estimated value differences of intermediate outcomes, we can achieve a smaller asymptotic variance. This result motivates us to construct the calibrated value estimator of $V(d) $ as
 \begin{equation}\label{value_CODA}
 \widehat{V}(d) = \widehat{V}_{P}(d) - \widehat{\boldsymbol{\rho}}(d)^\top \widehat{\Sigma}_M^{-1}(d) \{\widehat{W}_{P}(d)  - \widehat{W}_{U}(d)  \},
\end{equation}
where $ \widehat{\boldsymbol{\rho}}(d) $ is the estimator for $\boldsymbol{\rho}(d)$, and $\widehat{\Sigma}_M(d)$ is the estimator for $\Sigma_M(d)$. These variances can be consistently estimated by a simple plug-in method without accounting for the variation in estimating nuisance functions, such as propensity scores and conditional mean models of outcomes, due to the rate double robustness properties. See the detailed estimation in Section \ref{sec:est_Pho}. 
 Finally, the optimal decision rule under CODA for $X_P \sim X_U$, namely CODA-HO, is to maximize the calibrated value estimator within a pre-specified class of decision rules $\Pi$ as 
$\widehat{d} =\argmax_{d \in \Pi}  \widehat{V}(d),$ 
with the corresponding estimated value function as $\widehat{V}(\widehat{d})$. 



{\subsection{Calibrated Optimal Decision Making for Heterogeneous Baseline Covariates}\label{sec:CODA_hete}
We next consider a more challenging case where the distributions of baseline covariates in the primary sample and the auxiliary sample are distinct, $X_P\not \sim X_U$. The results under Lemma \ref{lem_cio} may not hold when the joint density of $X_P$ differs from the joint density of $X_U$. 
 As such, we need to construct a new estimator of modified value differences such that it converges to a normal distribution with a zero mean even under $X_P\not \sim X_U$.   
To this end, we consider rebalancing the value estimators of common intermediate outcomes in two samples based on their posterior sampling probability.} 
{Specifically, we combine two samples together and denote the joint dataset as 
$\{X_i, A_i, M_i, R_i, R_iY_i\}_{i=1,\cdots, n},\text{ for } n = N_P+N_U, $
where $R_i =1$ if subject $i$ is from the primary sample and $R_i =0$ if subject $i$ is from the auxiliary sample. Here, the distributions of baseline covariates, treatments, and intermediate outcomes are allowed to be different across different sub-samples, which distinguishes our work from the homogenous setting 
\citep[see][]{yang2019combining,athey2020combining,kallus2020role}.}

{To address the heterogeneous baseline covariates in two samples, also known as the covariate shift problem, a feasible way is to estimate the density functions of baseline covariates in two samples and adjust the corresponding estimator by the importance weights \citep[see][]{sugiyama2007covariate,kallus2021more}. However, these methods cannot handle a relatively large number of baseline covariates due to the estimation of density functions, and can be hard to develop a simple inference procedure. Instead, we use a similar projection approach as developed in Section \ref{sec:CODA_homo} to construct a new calibrated estimator through rebalancing  to handle the heterogeneous baseline covariates and gain efficiency.} To be specific, define the joint density of $\{X_i,A_i,M_i\}$ given $R_i$ as
 $f(X_i=x,A_i=a,M_i=m\mid R_i=1) \equiv  f_P(x,a,m)$ and $f(X_i=x,A_i=a,M_i=m\mid R_i=0) \equiv  f_U(x,a,m)$, respectively, where $f_P(x,a,m)$ and $f_U(x,a,m)$ are the joint density function of $\{X_P,A_P,M_P\}$ in the primary sample and the joint density function of $\{X_U,A_U,M_U\}$ in the auxiliary sample, respectively.
{By Bayesian theorem, the posterior sampling probability is 
\begin{equation}\label{sampling_prob}  
\prob(R_i=1\mid X_i=x,A_i=a,M_i=m) = {\prob(R_i=1) f_P(x,a,m) \over \prob(R_i=1) f_P(x,a,m) + \prob(R_i=0) f_U(x,a,m)}.
 \end{equation}
Here, we have $
\prob(R_i=1) = \lim_{N_P\to \infty} N_P/(N_P+N_U) = \lim_{N_P\to \infty} t/(1+t) = T/(1+T).$ 
Based on \eqref{sampling_prob}, we can rebalance the value estimators of common intermediate outcomes in each sample to construct a new mean zero estimator. To this end, we estimate the posterior sampling probability $r(x,a,m) \equiv \prob(R_i=1\mid X_i=x,A_i=a,M_i=m)$. Let $\widehat{r}(x,a,m)$ denote the resulting estimator. In addition, we estimate the propensity score function $\prob(A_i=1\mid X_i)$ in the joint sample, denoted as $\widehat{\pi}(X_i)$. For each sub-sample, we have new DR estimators for intermediate outcomes that taking the sampling probabilities into account as
\begin{equation*} 
 \begin{split}
\widehat{W}_{1}(d)=   {1\over n}\sum_{i=1}^{n} {R_i\over\widehat{r}\{X_i,d(X_i),M_i\} }{\mathbb{I}\{A_{i}=d(X_{i})\} [M_{i}  - \widehat{\theta} \{X_{i},d(X_{i})\} ]\over{A_{i} \widehat{\pi}(X_{i})+(1-A_{i})\{1-\widehat{\pi}(X_{i})\}}} + \widehat{\theta} \{X_{i},d(X_{i})\} ,\\
\widehat{W}_{0}(d)=   {1\over n}\sum_{i=1}^{n} {(1-R_i)\over1- \widehat{r}\{X_i,d(X_i),M_i\} }{\mathbb{I}\{A_{i}=d(X_{i})\}[M_{i}  - \widehat{\theta} \{X_{i},d(X_{i})\} ]\over{A_{i} \widehat{\pi}(X_{i})+(1-A_{i})\{1-\widehat{\pi}(X_{i})\}}}  + \widehat{\theta} \{X_{i},d(X_{i})\} ,
\end{split}
\end{equation*} 
where $\widehat{W}_{1}(d)$ and $\widehat{W}_{0}(d)$ are $s\times 1$ vectors, and $\widehat{\theta}(x,a)$ is the estimated conditional mean function for $\theta(x,a)$ as used in Section  \ref{sec:CODA_homo}. It can be shown that both $\widehat{W}_{1}(d)$ and $\widehat{W}_{0}(d)$ have asymptotic normality and converge to the same mean, 
as stated in the following lemma, under some technical conditions (A6) and (A7) detailed in Section \ref{sec:thms}. 
\begin{lemma}\label{lem2_hete}
Assume conditions (A1)-(A6) and (A7. i, iv, and v) hold. 
We have
\begin{equation*}
\begin{split}
\sqrt{n} \Big\{\widehat{W}_{1}(d)  -  W^*(d) \Big\} \rightsquigarrow N_s\Big\{\boldsymbol{0}_{s}, \Sigma_{1}(d)\Big\},   \text{ and }
\sqrt{n} \Big\{\widehat{W}_{0}(d)  -  W^*(d) \Big\} \rightsquigarrow N_s\Big\{\boldsymbol{0}_{s}, \Sigma_{0}(d)\Big\},
\end{split}
\end{equation*}
where $\Sigma_{1}$ and $\Sigma_{0}$ are $s\times s$ asymptotic covariance matrices for each sub-sample, and 
$W^*(d) = \int E\{M\mid d(X),X\}\{\prob(R=1) f_P(X) + \prob(R=0) f_U(X)\} dX,$ 
where $ f_P(X)$ is the marginal density of baseline covariates in the primary sample and $f_U(X)$ is the marginal density of baseline covariates in the auxiliary sample.
\end{lemma}
Here, to show Lemma \ref{lem2_hete}, we only require $E(M\mid X,A, R=1) = E(M\mid X,A,R=0)$, as indicated by assumption (A4). This is checkable since $(X,A, M)$ are observed in both samples. In contrast,  current methods handling multiple datasets \citep[see][]{kallus2020role,athey2020combining} require the missing at random assumption such that $R$ is independent of $Y$ given $(X,A,M)$. This implies $E(Y\mid X,A,M, R=1) = E(Y\mid X,A,M,R=0)$. This assumption is not testable due to the unobserved outcome in the auxiliary sample. Therefore, our method is built upon a more practical assumption.} The proof of Lemma \ref{lem2_hete} can be found in the Appendix. 
 {It is immediate from Lemma \ref{lem2_hete} that 
 $\widehat{W}_{1}(d) - \widehat{W}_{0}(d)$ is a mean zero estimator. The following lemma establishes the asymptotic normality of the new estimator. 
 \begin{lemma}\label{lem3_hete}
Suppose the conditions in Lemma \ref{lem2_hete} hold. 
We have
\begin{equation*}
\sqrt{n} \{\widehat{W}_{1}(d) - \widehat{W}_{0}(d)\}   \rightsquigarrow  N_s\Big\{\boldsymbol{0}_{s}, \Sigma_{R}(d)\Big\},
\end{equation*}
where $\Sigma_{R}(d) $ is a $s\times s$ asymptotic covariance matrix.
\end{lemma} }

   
       
 {
 Based on the results in \eqref{lem1} and Lemma \ref{lem3_hete}, the asymptotic joint distribution is
 \begin{equation*}\label{eqn:joint_hete}  
  \begin{split}
&\sqrt{N_P}		 \begin{bmatrix}
   			\widehat{V}_{P}(d)  -  V(d) \\
   			\sqrt{{n / N_P}}\{\widehat{W}_{1}(d) - \widehat{W}_{0}(d)\} \end{bmatrix}
	 \rightsquigarrow N_{s+1} \Bigg\{\boldsymbol{0}_{s+1}, \begin{bmatrix}
   			\sigma_Y^2(d) , \boldsymbol{\rho}_{R}(d)^\top\\
   			\boldsymbol{\rho}_{R}(d), \Sigma_{R}(d)
			\end{bmatrix}\Bigg\},    \text{ for all } d(\cdot),
\end{split}
\end{equation*}
where $\boldsymbol{\rho}_{R}(d)$ is the $s\times 1$ asymptotic covariance vector between the value estimator of the outcome of interest in the primary sample and the new rebalanced value difference estimator of intermediate outcomes between two samples. 
Following similar arguments in \eqref{eqn:proj}, it yields the calibrated value estimator of $V(d) $ under heterogeneous baseline covariates 
 \begin{equation}\label{value_CODA_hete}
 \widehat{V}_{R}(d) = \widehat{V}_{P}(d) - \sqrt{n/N_P} \widehat{\boldsymbol{\rho}}_{R}(d)^\top \widehat{\Sigma}_{R}^{-1}(d)  \{\widehat{W}_{1}(d) - \widehat{W}_{0}(d)\},
\end{equation}
where $ \widehat{\boldsymbol{\rho}}_{R}(d) $ is the estimator for $\boldsymbol{\rho}_{R}(d)$, and $\widehat{\Sigma}_{R}(d)$ is the estimator for $\Sigma_{R}(d)$.  Estimation on variances can be easily obtained using the simple plug-in method due to the rate double robustness as provided in Section \ref{sec:est_Pho}. 
Under the homogenous case where $f_P(x,a,m) = f_U(x,a,m)$, according to \eqref{sampling_prob}, we have  $\prob(R_i=1\mid X_i=x,A_i=a,M_i=m) = \prob(R_i=1)$. Then the above projection estimator will reduce to the estimator in Section \ref{sec:CODA_homo}.} {Therefore, ODR under CODA for $X_P\not \sim X_U$, namely CODA-HE, is to maximize the new calibrated value estimator $ \widehat{V}_{R}(d)$ within a pre-specified class of decision rules $\Pi$ as 
$\widehat{d}_{R} =\argmax_{d \in \Pi}   \widehat{V}_{R}(d),$ 
with the corresponding estimated value function as $\widehat{V}_{R}(\widehat{d}_{R})$.} 

 \subsection{Estimation on Variances}\label{sec:est_Pho}
{In this section, we present the estimators for ${\sigma}_Y^2 $, $\boldsymbol{\rho}$, $\boldsymbol{\rho}_{R}$, $\Sigma_M $, and $\Sigma_{R} $.  }
Recall $\widehat{\pi}_P$, $\widehat{\pi}_U$, {$\widehat{\pi}$, $ \widehat{\mu}_P $, $\widehat{\theta}$, and $\widehat{ r}$ are estimators for the propensity score functions $ \pi_P $, $\pi_U  $, and $\pi$, the conditional mean functions $ \mu_P$ and $\theta$, and the posterior sampling probability $ r $}, respectively, using any parametric or nonparametric models such as Random Forest or Deep Learning. Our theoretical results still hold with these nonparametric estimators as long as the estimators have desired convergence rates (see results established in \citet{wager2018estimation,farrell2021deep}).  To introduce the variance estimators, we first define the value functions at the individual level. Given a decision rule $d(\cdot)$, let the value for the $i$-th individual in terms of the outcome of interest as 
 \begin{equation*}
\widehat{v}^{(i)}_P(d) \equiv   {\mathbb{I}\{A_{P,i}=d(X_{P,i})\} [Y_{P,i} - \widehat{\mu}_P\{X_{P,i},d(X_{P,i})\} ] \over{A_{P,i} \widehat{\pi}_P (X_{P,i})+(1-A_{P,i})\{1-\widehat{\pi}_P(X_{P,i})\}}}   + \widehat{\mu}_P\{X_{P,i},d(X_{P,i})\}, 
 \end{equation*}
in the primary sample, for $i \in \{1,\cdots ,N_P\}$.  Similarly, the value for the $i$-th individual in terms of intermediate outcomes are
 \begin{equation*}
\widehat{\boldsymbol{w}}^{(i)}_P(d) \equiv   {\mathbb{I}\{A_{P,i}=d(X_{P,i})\} [M_{P,i}  - \widehat{\theta} \{X_{P,i},d(X_{P,i})\} ]\over{A_{P,i} \widehat{\pi}_P (X_{P,i})+(1-A_{P,i})\{1-\widehat{\pi}_P(X_{P,i})\}}}   + \widehat{\theta} \{X_{P,i},d(X_{P,i})\} ,
 \end{equation*}
in the primary sample for $i \in \{1,\cdots ,N_P\}$, and  
 \begin{equation*}
\widehat{\boldsymbol{w}}^{(i)}_U(d) \equiv  {\mathbb{I}\{A_{U,i}=d(X_{U,i})\} [M_{U,i}  - \widehat{\theta} \{X_{U,i},d(X_{U,i})\} ]\over{A_{U,i} \widehat{\pi}_U (X_{U,i})+(1-A_{U,i})\{1-\widehat{\pi}_U(X_{U,i})\}}}   + \widehat{\theta} \{X_{U,i},d(X_{U,i})\}, 
 \end{equation*}
in the auxiliary sample for $i \in \{1,\cdots ,N_U\}$, where $\widehat{\boldsymbol{w}}^{(i)}_P(d)$ and $\widehat{\boldsymbol{w}}^{(i)}_U(d)$ are $s\times 1$ vectors.
Following the results in \eqref{lem1} and Lemma \ref{lem3}, we propose to estimate  ${\sigma}_Y^2 (\cdot)$, $\boldsymbol{\rho}(\cdot)$ and $\Sigma_M (\cdot)$ by
\begin{eqnarray}\label{v_sd_est}
&& \widehat{\sigma}_Y^2 (d)={1\over N_P}\sum_{i=1}^{N_P}\{\widehat{v}_P^{(i)}(d) - \widehat{V}_P(d)\}^2,\\\nonumber
   &&\widehat{\boldsymbol{\rho}}(d)  = {1\over N_P}\sum_{i=1}^{N_P} \Big\{\widehat{v}_P^{(i)}(d) -\widehat{V}_P(d) \Big\} \Big\{\widehat{\boldsymbol{w}}^{(i)}_P(d) - \widehat{W}_P(d) \Big\},  \\\nonumber
  && \widehat{\Sigma}_M(d) ={1\over N_P}\sum_{i=1}^{N_P} \Big\{\widehat{\boldsymbol{w}}^{(i)}_P(d) -\widehat{W}_P(d) \Big\}^{\otimes 2} + t {1\over N_U}\sum_{i=1}^{N_U} \Big\{\widehat{\boldsymbol{w}}^{(i)}_U(d) -\widehat{W}_U(d) \Big\}^{\otimes 2},
\end{eqnarray}  
where $\eta^{\otimes 2} \equiv  \eta\eta^\top$ for $\eta$ as a vector.
{Similarly, based on the results in Lemma \ref{lem3_hete}, we define the rebalanced value for the $i$-th individual in terms of intermediate outcomes as
 \begin{equation*}
 \begin{split}
\widehat{\boldsymbol{w}}^{(i)}_1(d) \equiv   {R_i\over\widehat{r}\{X_i,d(X_i),M_i\} }{\mathbb{I}\{A_{i}=d(X_{i})\}[M_{i}  - \widehat{\theta} \{X_{i},d(X_{i})\} ]\over{A_{i} \widehat{\pi}(X_{i})+(1-A_{i})\{1-\widehat{\pi}(X_{i})\}}}  + \widehat{\theta} \{X_{i},d(X_{i})\},
 \end{split}
 \end{equation*}
 \begin{equation*}
  \begin{split}
\widehat{\boldsymbol{w}}^{(i)}_0(d) \equiv  {(1-R_i)\over1- \widehat{r}\{X_i,d(X_i),M_i\} }{\mathbb{I}\{A_{i}=d(X_{i})\} [M_{i}  - \widehat{\theta} \{X_{i},d(X_{i})\} ]\over{A_{i} \widehat{\pi}(X_{i})+(1-A_{i})\{1-\widehat{\pi}(X_{i})\}}} + \widehat{\theta} \{X_{i},d(X_{i})\},
 \end{split}
 \end{equation*}
 where $\widehat{\boldsymbol{w}}^{(i)}_0(d)$ and $\widehat{\boldsymbol{w}}^{(i)}_1(d)$ are $s\times 1$ vectors, for $i=1,\cdots,  N_P+N_U$. Also, we define the correlated part for the $i$-th individual in terms of intermediate outcomes as 
  \begin{equation*} 
\widehat{\boldsymbol{\psi}}^{(i)}(d) \equiv   {1\over\widehat{r}\{X_i,d(X_i),M_i\} }{\mathbb{I}\{A_{i}=d(X_{i})\}\over{A_{i} \widehat{\pi}(X_{i})+(1-A_{i})\{1-\widehat{\pi}(X_{i})\}}}  [M_{i}  - \widehat{\theta} \{X_{i},d(X_{i})\} ],
 \end{equation*}
 where $\widehat{\boldsymbol{\psi}}^{(i)}(d)$ is $s\times 1$ vectors, for $i=1,\cdots, N_P$. We then propose to estimate $\boldsymbol{\rho}_{R}(\cdot)$ and $\Sigma_{R} (\cdot)$ by
\begin{eqnarray}\label{v_sd_est_hete} 
  \widehat{\boldsymbol{\rho}}_{R} (d)  = && {1\over N_P}\sum_{i=1}^{N_P}  \Big\{\widehat{v}_P^{(i)}(d) -\widehat{V}_P(d) \Big\} \sqrt{N_P/n} \widehat{\boldsymbol{\psi}}^{(i)}(d) ,  \\\nonumber
  \widehat{\Sigma}_{R} (d) = && {1\over n}\sum_{i=1}^{n}  \Big[ \{\widehat{\boldsymbol{w}}^{(i)}_1(d) -\widehat{\boldsymbol{w}}^{(i)}_0(d)\}\Big]^{\otimes 2}.
\end{eqnarray}  }

\subsection{Iterative Policy Tree Search Algorithm}\label{sec:impl}
We introduce the iterative policy tree search algorithm to implement CODA. We first elaborate on how to find the ODR that maximizes the calibrated value estimator for the homogenous case. 
Following the tree-based policy learning algorithm proposed in \cite{athey2017efficient}, we define the reward of the $i$-th individual in the primary sample by $\widehat{v}^{(i)}_P(d) $ 
for $i \in \{1,\cdots ,N_P\}$.  Specifically, the reward of the $i$-th individual is $\widehat{v}^{(i)}_P(1)$ under treatment 1 and  $\widehat{v}^{(i)}_P(0)$ under treatment 0. The decision tree allocates individuals to different treatments, and receives the corresponding rewards. The ODR based on the primary sample solely is obtained by maximizing the sum of these rewards through the exhaustive search within $\Pi_1$, denoted as
$\widehat{d}_P =\underset{d\in \Pi_1}{\argmax}  \sum_{i=1}^{N_P} \widehat{v}_P^{(i)} (d).$ 
To develop ODR by CODA-HO within the class $\Pi_1$, from \eqref{value_CODA}, we construct the calibrated reward of the $i$-th individual in the primary sample by  
 \begin{equation}\label{tree_CODA_reward}
\widehat{v}^{(i)} (d) =\widehat{v}^{(i)}_P(d)- \widehat{\boldsymbol{\rho}}(d)^\top \widehat{\Sigma}_M^{-1}(d) \{\widehat{\boldsymbol{w}}^{(i)}_P(d) - \widehat{W}_{U}(d)  \}.
\end{equation}
Here, notice that the sample mean of \eqref{tree_CODA_reward} over the primary sample yields \eqref{value_CODA}.
Therefore, the decision tree that maximizes the sum of rewards defined in \eqref{tree_CODA_reward} also maximizes the calibrated value estimator in \eqref{value_CODA}. The finite-depth tree-based ODR under CODA-HO is 
 \begin{equation}\label{eqn:CODA_tree}
\widehat{d} =\underset{d\in \Pi_1}{\argmax}  \sum_{i=1}^{N_P} \widehat{v}^{(i)} (d).
\end{equation} 
Yet, the estimators $\widehat{\boldsymbol{\rho}}(d)$ and $ \widehat{\Sigma}_M^{-1}(d)$ defined in \eqref{tree_CODA_reward} are calculated using two samples' information (see details in Section \ref{sec:est_Pho}) 
given a specific decision rule $d$, and thus the tree-based policy learning in \cite{athey2017efficient} is not directly applicable to solve \eqref{eqn:CODA_tree}. 
To address this difficulty, 
we propose an {iterative policy tree search algorithm} consisting of four steps as follows.

\noindent \textbf{Step 1.} Find the ODR based on the primary sample solely ($\widehat{d}_P$) as an initial tree.

\noindent \textbf{Step 2.} Estimate $\boldsymbol{\rho}(\cdot)$ and $\Sigma_M (\cdot)$ 
by plugging in $d =\widehat{d}_P$.  Thus, the calibrated reward for the $i$-th individual can be approximated by 
  $\widehat{v}^{(i)}_P(1)- \widehat{\boldsymbol{\rho}}(\widehat{d}_P)^\top \widehat{\Sigma}_M^{-1}(\widehat{d}_P) \{\widehat{\boldsymbol{w}}^{(i)}_P(1) - \widehat{W}_{U}(1)  \}$
    under treatment 1, and  
  $ \widehat{v}^{(i)}_P(0)- \widehat{\boldsymbol{\rho}}(\widehat{d}_P)^\top \widehat{\Sigma}_M^{-1}(\widehat{d}_P) \{\widehat{\boldsymbol{w}}^{(i)}_P(0) - \widehat{W}_{U}(0)  \} $
   under treatment 0.
   
   \noindent \textbf{Step 3.} Search for the optimal decision tree within the class $\Pi_1$ to achieve a maximum overall calibrated reward, denoted as $\widehat{d}^{(1)}$. This step can be solved by applying the tree-based policy learning in \cite{athey2017efficient} with an updated reward matrix.

   \noindent \textbf{Step 4.}  Repeat steps 2 and 3 for $k=1,\cdots,K$, by replacing the previous estimated decision tree $\widehat{d}^{(k-1)}$ ($\widehat{d}^{(0)}=\widehat{d}_P$) with the new estimated decision tree $\widehat{d}^{(k)}$ until it's convergent or achieves the maximum number of iterations $K$. It is observed in Section \ref{sec:4} that $\widehat{d}_P$ is fairly close to $\widehat{d}$, and thus one iteration is usually sufficient to find ODR under CODA in practice.  
    
 
The above iterative policy search algorithm can be extended to the heterogeneous case and parametric decision rules, as provided in Section \ref{sec:ext_algo}.

\subsubsection{Extension of Iterative Policy Tree Search Algorithm}\label{sec:ext_algo}

 {
 
 The iterative policy search algorithm also works for finding ODR under CODA-HE, 
 by replacing the corresponding variance estimates from \eqref{v_sd_est} with that from \eqref{v_sd_est_hete}. Specifically, we can rewrite the calibrated value estimator in \eqref{value_CODA_hete} as
  \begin{eqnarray*}
 \widehat{V}_{R}(d) 
 =&{1\over N_P}\sum_{i=1}^{N_P} \widehat{v}_P^{(i)}(d) - \sqrt{{n\over N_P}} \widehat{\boldsymbol{\rho}}_{R}(d)^\top \widehat{\Sigma}_{R}^{-1}(d)   {1\over n}\sum_{i=1}^{n}\{\widehat{\boldsymbol{w}}^{(i)}_1(d) - \widehat{\boldsymbol{w}}^{(i)}_0(d)\}\\\nonumber
 =&{1\over N_P}\sum_{i=1}^{N_P} \widehat{v}_P^{(i)}(d) - \sqrt{{n\over N_P}}\widehat{\boldsymbol{\rho}}_{R}(d)^\top \widehat{\Sigma}_{R}^{-1}(d)  \left[ {1\over N_P}\sum_{i=1}^{N_P} {N_P\over n}\{\widehat{\boldsymbol{w}}^{(i)}_1(d)   - \widehat{\boldsymbol{w}}^{(i)}_0(d)\}\right.\\\nonumber
 &\left.+{1\over n}\sum_{i=N_P+1}^{n} \{\widehat{\boldsymbol{w}}^{(i)}_1(d)   - \widehat{\boldsymbol{w}}^{(i)}_0(d)\}\right]. 
\end{eqnarray*}
This motivates the calibrated reward for the $i$-th individual in the heterogeneous case as
 \begin{equation*}  
  \widehat{v}^{(i)}_P(1)- \sqrt{{n\over N_P}} \widehat{\boldsymbol{\rho}}_{R}(\widehat{d}_P)^\top \widehat{\Sigma}_{R}^{-1}(\widehat{d}_P)  \left[   {N_P\over n}\{\widehat{\boldsymbol{w}}^{(i)}_1(1)   - \widehat{\boldsymbol{w}}^{(i)}_0(1)\} +\widehat{\Delta}(1)\right],
   \end{equation*}  
    under treatment 1, where $\widehat{\Delta}(1) = {n^{-1}}\sum_{i=N_P+1}^{n} \{\widehat{\boldsymbol{w}}^{(i)}_1(1)   - \widehat{\boldsymbol{w}}^{(i)}_0(1)\}$, and  
 \begin{equation*}  
  \widehat{v}^{(i)}_P(0)- \sqrt{{n\over N_P}} \widehat{\boldsymbol{\rho}}_{R}(\widehat{d}_P)^\top \widehat{\Sigma}_{R}^{-1}(\widehat{d}_P)  \left[   {N_P\over n}\{\widehat{\boldsymbol{w}}^{(i)}_1(0)   - \widehat{\boldsymbol{w}}^{(i)}_0(0)\} +\widehat{\Delta}(0)\right],
   \end{equation*} 
   under treatment 0, where $\widehat{\Delta}(0) = {n^{-1}}\sum_{i=N_P+1}^{n} \{\widehat{\boldsymbol{w}}^{(i)}_1(0)   - \widehat{\boldsymbol{w}}^{(i)}_0(0)\}$. Then, using similar steps in the iterative policy tree search algorithm for CODA-HO, we can find ODR under CODA-HE.
 }

We next extent the iterative policy tree search to parametric decision rules with an illustration in homogenous case. Suppose the decision rule $d(\cdot)$ relies on a model parameter $\beta$, denoted as $d(\cdot)\equiv d(\cdot;\beta)$. We use a shorthand to write $V(d)$ as $V(\beta)$, and define
$
\beta_0 =\argmax_{\beta} V(\beta).
$
Thus, the value for the primary outcome of interest under the true ODR $d(\cdot;\beta_0)$ is defined as $V(\beta_0)$. Suppose the decision rule takes a form as $d(X;\beta) \equiv \mathbb{I}\{g(X)^\top  \beta > 0\}$, where $g(\cdot)$ is an unknown function and $\mathbb{I}(\cdot)$ is the indicator function. We use $\phi_X(\cdot)$ to denote a set of basis functions of baseline covariates with length $q$, which are rich enough to approximate the underlying function $g(\cdot)$. Thus, the decision rule is found within a class of $\mathbb{I}\{\phi_X(X)^\top  \beta > 0\}$, denoted as the class $\Pi_2$. Here, for notational simplicity, we include 1 in $\phi_X(\cdot)$ so that the parameter $\beta \in  \mathbb{R}^{q+1}$.

Suppose the decision rule $d(\cdot)$ falls in class $\Pi_2$ that relies on a parametric model with parameters $\beta$. We use shorthands to write $\boldsymbol{\rho}(d)$ as $\boldsymbol{\rho}(\beta)$, $\Sigma_M(d)$ as $\Sigma_M(\beta)$, $\widehat{V}_{P}(d)$ as $\widehat{V}_{P}(\beta)$, $\widehat{W}_{P}(d)$ as $\widehat{W}_{P}(\beta)$, and $\widehat{W}_{U}(d)$ as $\widehat{W}_{U}(\beta)$, respectively. Then, the  calibrated value estimator for $V(\beta) $ can be constructed by 
\begin{equation*}  
 \widehat{V}(\beta) = \widehat{V}_{P}(\beta) + \widehat{\boldsymbol{\rho}}(\beta)^\top \widehat{\Sigma}_M^{-1}(\beta) \{\widehat{W}_{P}(\beta)  - \widehat{W}_{U}(\beta)  \},
\end{equation*}
where $ \widehat{\boldsymbol{\rho}}(\beta) $ is the estimator for $\boldsymbol{\rho}(\beta)$, and $\widehat{\Sigma}_M(\beta)$ is the estimator for $\Sigma_M(\beta)$. 
We apply a similar iterative-updating procedure discussed above for CODA-HO but within parametric decision rules. 

\noindent \textbf{Step 1*.} Find the ODR in the primary sample by $\widehat{\beta}_P = {\argmax}_\beta  \widehat{V}_P(\beta)$ as an initial decision rule. This can be solved using any global optimization algorithm, such as the heuristic algorithm provided in the \proglang{R} package \code{rgenound}.

\noindent \textbf{Step 2*.}  Estimate covariances $\boldsymbol{\rho}(\beta)$ and $\Sigma_M (\beta)$ based on \eqref{v_sd_est} with $\beta = \widehat{\beta}_P$.   
Then, we search the ODR based on the calibrated value estimator within the class $\Pi_2$ that maximizes
 \begin{equation*}  
 \widehat{V}(\beta) = \widehat{V}_{P}(\beta) - \widehat{\boldsymbol{\rho}}(\widehat{\beta}_P)^\top \widehat{\Sigma}_M^{-1}(\widehat{\beta}_P) \{\widehat{W}_{P}(\beta)  - \widehat{W}_{U}(\beta)  \}.
\end{equation*}  

\noindent \textbf{Step 3*.} Repeat step 2* for $k=1,\cdots,K$, by replacing the previous estimated $\widehat{\beta}^{(k-1)}$ ($\widehat{\beta}^{(0)}=\widehat{\beta}_P$) with the new estimated $\widehat{\beta}^{(k)}$ until the number of iterations achieves $K$ or $||\widehat{\beta}^{(k)}- \widehat{\beta}^{(k-1)}||_2 <\delta$ where $\delta$ is a pre-specified threshold and $||\cdot||_2 $ is the $L_2$ norm.

\section{Theoretical Properties}\label{sec:thms}
 {In this section, we investigate the theoretical properties of the value estimator under CODA in \eqref{value_CODA} and \eqref{value_CODA_hete}, 
 respectively. 
 All the proofs are provided in the Appendix.} 
 As standard in personalized decision-making \citep{zhang2012robust,luedtke2016statistical,kitagawa2018should,rai2018statistical}, we introduce the following technical conditions.
 



\noindent \textbf{(A6)}. Suppose the supports $\mathbb{X}_P$, $\mathbb{X}_U$, $\mathbb{M}_P$, $\mathbb{M}_U$, and $\mathbb{Y}_P$ are bounded.


%
\noindent \textbf{(A7)}. Rate double robustness for model misspecification: for $ a = 0,1,$ 
\begin{eqnarray*} 
&& \text{ (i) } [\Mean_{X \in \mathbb{X}_P} \{\mu_P(X,a)-\widehat{\mu}_P(X,a)\}^2 \{\pi_P(X)-\widehat{\pi}_P(X)\}^2]^{1\over2} =o_p(N_P^{-1/ 2}),\\
 &&\text{ (ii),  (iii) } [\Mean_{X \in \mathbb{X}_J} \{\theta(X,a)-\widehat{\theta}(X,a)\}^2 \{\pi_J(X)-\widehat{\pi}_J(X)\}^2]^{1\over2} =o_p(N_J^{-1/ 2}), \quad J\in \{P,U\},\\
 &&\text{ (iv) } [\Mean_{X \in \mathbb{X}_P\cup \mathbb{X}_U} [\{\theta(X,a)-\widehat{\theta}(X,a)\}^2  \{\pi(X)-\widehat{\pi}(X)\}^2]^{1\over2} =o_p(n^{-1/ 2}),\\
 &&\text{ (v) }  [ \Mean_{X \in \mathbb{X}_P\cup \mathbb{X}_U, M \in \mathbb{M}_P\cup \mathbb{M}_U} [\{\theta(X,a)-\widehat{\theta}(X,a)\}^2\{r(X,a,M)-\widehat{r}(X,a,M)\}^2]^{1\over2} =o_p(n^{-1/ 2}).
\end{eqnarray*}  

\noindent \textbf{(A8)}. There exist some constants $\gamma,\lambda>0$ such that $ \prob\{0<|\Mean(Y_P\mid X_P,A_P=1)-\Mean(Y_P\mid X_P,A_P=0)|\le \xi \}=O(\xi^{\gamma}),$   where the big-$O$ term is uniform in $0<\xi \le \lambda$.

Assumption (A6) is a technical assumption sufficient to establish the uniform convergence results
in the literature of optimal treatment regime estimation \citep[see][]{zhang2012robust,zhao2012estimating,zhou2017residual}. Assumption (A7) (i)-(iv) requires the estimated conditional mean outcomes and propensity score functions to converge at certain rates in each decision-making problem. 
This assumption is commonly imposed in the causal inference literature \citep[see][]{athey2020combining, kallus2020role} to derive the asymptotic distribution of the estimated average treatment effect with either parametric or non-parametric estimators \citep[see][]{wager2018estimation,farrell2021deep}. {We extend it to (A7) (v) that the estimators of the posterior sampling probability and conditional mean functions of intermediate outcomes converge at certain rates, by which together with (A7) (iv) one can establish the asymptotic distribution of the rebalanced value estimators based on the joint sample.}
Finally, assumption (A8) is well known as the margin condition, which is often adopted in the literature to derive a sharp convergence rate for the value function under the estimated optimal decision rule \citep[see][]{qian2011performance,luedtke2016statistical,kitagawa2018should}. 
%
%
%
%
%
{We first establish the consistency of the proposed estimators in the following theorem.
\begin{theorem}\label{thm2}
Suppose conditions (A1)-(A7) hold. 
We have \\
(i)  $\widehat{\sigma}_Y^2 (d) ={\sigma}_Y^2 (d) +o_p(1);$ (ii) $\widehat{\boldsymbol{\rho}}(d) =\boldsymbol{\rho}(d)+o_p(1);$ (iii) $\widehat{\Sigma}_M(d) =\Sigma_M(d)+o_p(1);$
\\
(iv) $\widehat{\boldsymbol{\rho}}_{R}(d) =\boldsymbol{\rho}_{R}(d)+o_p(1);$ (v) $\widehat{\Sigma}_{R}(d) =\Sigma_{R}(d)+o_p(1);$
\\
(vi) if $X_P \sim X_U$, $\widehat{V}(d) =V(d)+o_p(1);$ 
(vii) $\widehat{V}_{R}(d) =V(d)+o_p(1).$
\end{theorem}}
The result (vi) in Theorem \ref{thm2} requires additional homogeneous baseline covariates, while the rest results hold for either the homogeneous or the heterogeneous case. We remark that the key step of the proof is to decompose the variance estimators based on the true value estimator defined at the individual level in each sample. This allows replacing the estimators with their true models based on the rate doubly robustness in assumption (A7) with a small order. 
We next show the asymptotic normality of the proposed value under $X_P \sim X_U$.
{\begin{theorem}\label{thm3}
Suppose $ \{d^{opt}, \widehat{d}\}\in \Pi_1$ or $ \{d^{opt}, \widehat{d}\} \in \Pi_2$, (A1)-(A6), (A7. i, ii, and iii), and (A8) hold. Under $X_P \sim X_U$, let $\sigma^2(d^{opt})  = \sigma_Y^2(d^{opt}) - \boldsymbol{\rho}(d^{opt})^\top\Sigma_M^{-1}(d^{opt}) \boldsymbol{\rho}(d^{opt}) $, then 
 \begin{equation*}  
\sqrt{N_P} \{\widehat{V}(\widehat{d})-V(d^{opt}) \}   \rightsquigarrow N \{ 0, \sigma^2(d^{opt})   \}.
\end{equation*}
\end{theorem} }
The condition in Theorem \ref{thm3} that $ \{d^{opt}, \widehat{d} \}\in \Pi_1$ or $ \{d^{opt}, \widehat{d}\} \in \Pi_2$ requires the true ODR falls into the class of decision rules of interest, such that the resulting estimated decision rule by CODA-HO is not far away from the true ODR. The proof of Theorem \ref{thm3} consists of three steps. We first replace the estimated propensity score and conditional mean functions in $\widehat{V}(\widehat{d})$ by their counterparts based on (A7) with a small order $o_p(N_P^{-1/2})$. Secondly, we show the value estimator under the estimated decision rule by CODA-HO converges to the value estimator under the true ODR at a rate of $o_p(N_P^{-1/2})$, 
under (A8). The proof is nontrivial to the literature, by noticing that the estimated decision rule by CODA-HO is to maximize the newly proposed calibrated value estimator. Lastly, the asymptotic normality follows the central limit theorem. Since the two samples ($P$ and $U$) are independently collected from two separate studies, we can explicitly derive the asymptotic variance of the calibrated value estimator under the estimated decision rule by CODA-HO. 
By Theorem \ref{thm3}, when $\boldsymbol{\rho}(d^{opt})$ is a non-zero vector, we have $\sigma_Y^2 (d^{opt}) - \boldsymbol{\rho}(d^{opt})^\top\Sigma_M^{-1}(d^{opt})\boldsymbol{\rho}(d^{opt}) < \sigma_Y^2 (d^{opt})  $, since $\Sigma_M(d^{opt})$ is positive definite. In other words, if $Y$ is correlated with one of the selected $M$, the asymptotic variance of the calibrated value estimator under CODA-HO is strictly smaller than the asymptotic variance of the value estimator under the ODR obtained based on the primary sample solely. The larger the correlation is, the smaller variance we can achieve. Hence, the proposed calibrated value estimator is more efficient by integrating different data sources. 
Based on Theorem \ref{thm3}, By plugging the estimates $\widehat{\sigma}_Y^2 (\widehat{d})$, $\widehat{\boldsymbol{\rho}}(\widehat{d})$, and $\widehat{\Sigma}_M(\widehat{d})$, the asymptotic variance of $\widehat{V}(\widehat{d})$ can be consistently estimated by
$\widehat{\sigma}^2(\widehat{d})\equiv  \widehat{\sigma}_Y^2 (\widehat{d}) -\widehat{\boldsymbol{\rho}}(\widehat{d})^\top \widehat{\Sigma}_M^{-1}(\widehat{d}) \widehat{\boldsymbol{\rho}}(\widehat{d}).$ 
Therefore, a two-sided $1-\alpha$ confidence interval (CI) for $V(d^{opt})$ under CODA-HO 
is 
\begin{equation}\label{ciaipw}
\Big [ \widehat{V}(\widehat{d})-z_{\alpha/2}\widehat{\sigma}/\sqrt{N_P},\quad \widehat{V}(\widehat{d})+z_{\alpha/2}\widehat{\sigma}/\sqrt{N_P} \Big],
\end{equation}
where $z_{\alpha/2}$ denotes the upper $\alpha/2-$th quantile of a standard normal distribution. 
{Similarly, we establish the asymptotic normality for the heterogeneous case as follows.

\begin{theorem}\label{thm3_hete}
Suppose $ \{d^{opt}, \widehat{d}_{R}\}\in \Pi_1$ or $ \{d^{opt}, \widehat{d}_{R}\} \in \Pi_2$. Under (A1)-(A6), (A7. i, iv, and v), and (A8), define $\sigma_{R}^2(d^{opt})  = \sigma_Y^2(d^{opt}) - \boldsymbol{\rho}_{R}(d^{opt})^\top\Sigma_{R}^{-1}(d^{opt}) \boldsymbol{\rho}_{R}(d^{opt}) $, then 
 \begin{equation*}  
\sqrt{N_P} \{\widehat{V}_{R}(\widehat{d}_R)-V(d^{opt}) \}   \rightsquigarrow N \{ 0, \sigma_{R}^2(d^{opt})   \}.
\end{equation*}
\end{theorem} 
Here, we also require the true ODR falls into the class of decision rules of interest, such that the resulting estimated ODR by CODA-HE is not far away from the true. The proof of Theorem \ref{thm3_hete} is similar to Theorem \ref{thm3}. When $\boldsymbol{\rho}_R(d^{opt})$ is a non-zero vector, we have $\sigma_Y^2 (d^{opt}) - \boldsymbol{\rho}_R(d^{opt})^\top\Sigma_R^{-1}(d^{opt})\boldsymbol{\rho}_R(d^{opt}) < \sigma_Y^2 (d^{opt})  $ such that the proposed calibrated value estimator is more efficient. The corresponding two-sided $1-\alpha$ CI for $V(d^{opt})$ under CODA-HE is 
\begin{equation}\label{ciaipw_hete}
\Big [ \widehat{V}_R(\widehat{d}_R)-z_{\alpha/2}\widehat{\sigma}_R/\sqrt{N_P},\quad \widehat{V}_R(\widehat{d}_R)+z_{\alpha/2}\widehat{\sigma}_R/\sqrt{N_P} \Big],
\end{equation}
where $\widehat{\sigma}_R = \widehat{\sigma}_Y^2 (\widehat{d}_R) -\widehat{\boldsymbol{\rho}_R}(\widehat{d}_R)^\top \widehat{\Sigma}_R^{-1}(\widehat{d}_R) \widehat{\boldsymbol{\rho}_R}(\widehat{d}_R)$ is the estimator of $\sigma_{R}^2(d^{opt}) $. 
}

\section{Simulation Studies}\label{sec:4}


  \subsection{Evaluation on Calibrated Value Estimator for {Homogeneous Baseline Covariates}}\label{simu_eval}

Data are generated by: $A  {\sim} \text{Bernoulli} \{\pi(X)\}, ~~ X^{(1)}, \cdots,X^{(r)}  {\sim} \text{Uniform}[-2,2]$,~~ \\$M=U^M(X)+ A C^M(X)+\epsilon^J,~~ J\in \{P,U\}, ~~Y=U^Y(X)+AC^Y(X)+\epsilon^Y,$
where $U^M(\cdot)$ and $U^Y(\cdot)$ are the baseline functions of $M$ and $Y$, respectively, $C^M(\cdot)$ and  $C^Y(\cdot)$ are the contrast functions that describes the treatment-covariates interaction for $M$ and $Y$, respectively, and $\epsilon^J $ and $\epsilon^Y$ are the random errors. 
Let $\text{ logit}\{\pi(X)\} = 0.4+0.2X^{(1)} - 0.2X^{(2)}$. 
Consider two scenarios with $s=1,~ r=2$.

\noindent \textbf{Scenario 1}: $
		U^M(X)={X^{(1)}}+2X^{(2)},~~ C^M(X)=X^{(1)}  X^{(2)},~~ U^Y(X)=2{X^{(1)}}+X^{(2)}, ~~C^Y(X)=2{X^{(1)}}X^{(2)}.$

\noindent \textbf{Scenario 2}: $ 
		U^M(X)={X^{(1)}}+2X^{(2)},~~ C^M(X)=X^{(1)}-X^{(2)}, ~~U^Y(X)=2{X^{(1)}}+X^{(2)},\\~~ C^Y(X)=2\{{X^{(2)}}-X^{(1)}\}. $
	
	The noises of $M_U$ are set to be $\epsilon^{U} {\sim} \text{Uniform}[-1,1]$, while the noises of $M_P$ and the noise of $Y_P$ are generated from a bivariate normal distribution with mean zero, variance vector as $[2, 1.5]$, and a positive correlation of 0.7, to account for heterogeneity. 
For Scenario 1, the true ODR can be represented by a decision tree as $d^{opt}(X) = \mathbb{I}\{X^{(1)}{X^{(2)}}>0\}$, which is unique up to permutation. Its true value $V(d^{opt})$ can be calculated by Monte Carlo approximations, as 0.999. In Scenario 2, the true ODR takes a form of a linear rule as $d^{opt}(X) = \mathbb{I}\{{X^{(2)}}-X^{(1)}>0\}$, 
with its true value $V(d^{opt})$ as 1.333. Since $d^{opt}$ in Scenario 2 cannot be represented by a decision tree, $d^{opt}  \not \in \Pi_1$, we have $\max_{d\in \Pi_1}V(d) = 1.251$ for Scenario 2 as the true value under the optimal decision tree, which is smaller than $V(d^{opt})$.


We consider $N_U=2000$ and allow $N_P$ chosen from the set $\{500,1000\}$. 
we search the ODR using CODA-HO based on two samples, and the ODR based on the primary sample solely, within the class of decision trees $\Pi_1$. We denote the later method as `single baseline'. 
To illustrate the approximation error due to policy search, we directly plug the true decision rule $d^{opt}$ in the value estimators for comparison.  
The empirical results are reported in Table \ref{table:2}, 
 aggregated over 500 replications. We summarize the true value function $V(\cdot)$ of a given decision rule computed using Monte Carlo, the estimated value $\widehat{V}(\cdot)$ with its standard deviation $SD\{\widehat{V}(\cdot)\}$, the averaged estimated standard error $\Mean\{\widehat{\sigma}\}$, and the coverage probability based on the $95\%$  CI in \eqref{ciaipw}, for both CODA-HO and 
`single baseline', under the true ODR ($d^{opt}$) and the estimated rules ($\widehat{d}$ and $\widehat{d}_P$), respectively. In addition, we report the improved efficiency using CODA-HO, which is calculated as the relative reduction in the standard deviation of the CODA-HO value estimator with respect to that of `single baseline', 
the estimated asymptotic correlation $\widehat{\boldsymbol{\rho}}$, and the estimated asymptotic covariance $\widehat{\Sigma}_M$
for CODA-HO. 
 
 \begin{table}[!t]
\caption{Empirical results of the proposed CODA-HO method in comparison to the ODR based on the primary sample solely under Scenarios 1-2  with {homogeneous} baseline covariates, where $\max_{d\in \Pi_1}V(d) = 1.251$ in Scenario 2.}\label{table:2}
\begin{center}
\begin{tabular}{ccc|cc||cc|cc}
\hline\hline
Method (Rule)	&\multicolumn{2}{c|}{CODA ($d^{opt}$)} &\multicolumn{2}{c||}{CODA ($\widehat{d}$)} &\multicolumn{2}{c|}{ODR ($d^{opt}$)} &\multicolumn{2}{c}{ODR ($\widehat{d}_P$)} \\ \hline
$N_P=$	& $500$ & $1000$& $ 500$ & $ 1000$& $ 500$ & $ 1000$& $ 500$ & $ 1000$\\
\hline\hline
\textbf{Scenario 1}	&\\
\hline 
Plug-in Value  $V(\cdot)$ &\multicolumn{2}{c|}{0.999}  & 0.963   & 0.976   & \multicolumn{2}{c|}{0.999}   & 0.967   & 0.976    \\
\hline
 		\hline
Estimated Value $\widehat{V}(\cdot)$&0.998  & 0.996   & 1.053   & 1.030  & 1.006   & 0.994   & 1.095   & 1.050  \\
\hline
 $SD\{\widehat{V}(\cdot)\}$&0.124   & 0.098   & 0.123   & 0.098   & 0.173  & 0.127   & 0.171   & 0.125    \\
\hline
 $\Mean\{\widehat{\sigma}\}$&0.129   & 0.096   & 0.129   & 0.095   & 0.182   & 0.128   & 0.181   & 0.128    \\
\hline
Coverage Probabilities &95.6\% & 95.0\% & 94.6\% & 94.6\% & 96.8\% & 95.2\% & 94.2\% & 94.4\%  \\
\hline 
\hline
 Improved Efficiency &29.1\%   & 25.0\%  &28.7\%   & 25.8\%& /      &  /     &    /   &  /      \\
 \hline
 $\widehat{\boldsymbol{\rho}}(\cdot)$&12.4   & 12.3  &12.4  & 12.3   & /      &  /     &    /   &  /      \\
\hline 
 $\widehat{\Sigma}_M(\cdot)$ &18.8   & 20.8   & 18.8  & 20.8   &    /   &  /     &    /   &  /     \\
\hline\hline
\textbf{Scenario 2}	&\\
\hline 
Plug-in Value  $V(\cdot)$ &\multicolumn{2}{c|}{1.333}  & 1.236   & 1.239   & \multicolumn{2}{c|}{1.333}    & 1.227   & 1.232   \\
\hline
 		\hline
Estimated Value $\widehat{V}(\cdot)$&1.327   & 1.332   & 1.321   & 1.303  & 1.329   & 1.331   & 1.350   & 1.319    \\
\hline
 $SD\{\widehat{V}(\cdot)\}$&0.110   & 0.085  & 0.108   & 0.085  & 0.156   & 0.112   & 0.154 & 0.110\\
\hline
 $\Mean\{\widehat{\sigma}\}$&0.115   & 0.085   & 0.116   & 0.086   & 0.162   & 0.114  & 0.161   & 0.114    \\
\hline
Coverage Probabilities &95.8\% & 95.2\% & 96.2\% & 94.6\% & 95.4\% & 95.2\% & 96.4\% & 95.2\%  \\
\hline 
\hline
 Improved Efficiency &29.0\%   & 25.4\%  &28.0\%   & 24.6\%& /      &  /     &    /   &  /      \\
 \hline
 $\widehat{\boldsymbol{\rho}}(\cdot)$&10.7   & 10.6 &10.6  & 10.5   &     /  &  /     &     /  & /       \\
\hline 
 $\widehat{\Sigma}_M(\cdot)$ &17.8  & 19.4   & 17.8& 19.5& /      &/       &     /  & /       \\
\hline 
\end{tabular}
\end{center}
\end{table}

%
%


Based on Table \ref{table:2}, 
it is clear that CODA-HO is more efficient than `single baseline', 
 in all cases. 
 {To be specific, CODA-HO improves efficiency by 28.7\% in Scenario 1 and 28.0\% in Scenario 2 for $N_P = 500$, and by 25.8\% in Scenario 1 and 24.6\% in Scenario 2 for $N_P = 1000$. 
On the other hand, the values under CODA-HO approach the true as the sample size $N_P$ increases in all scenarios. Specifically, the proposed method achieves $V (\widehat{d}) = 0.977$ in Scenario 1 ($V (d^{opt}) = 0.999$) and $V (\widehat{d}) = 1.240$ in Scenario 2 $(\max_{d\in \Pi_1}V(d) = 1.251)$ when $N_P = 1000$. These results are comparable to or slightly better than the values under `single baseline'.} 
Two findings help to verify Theorem \ref{thm3}. First, the mean of the estimated standard error of the value function ($\Mean\{\widehat{\sigma}\}$) is close to the standard deviation of the estimated value ($SD\{\widehat{V}\}$), and gets smaller as the sample size $N_P$ increases. Second, the empirical coverage probabilities of the proposed 95\%  CI in \eqref{ciaipw} approach to the nominal level in all settings. All these findings are further justified by directly applying the true ODR into the proposed calibrated value estimator. 
It can be observed in Table \ref{table:2} 
that the estimated asymptotic correlation $\widehat{\boldsymbol{\rho}}$ and the estimated asymptotic covariance $\widehat{\Sigma}_M$ under the estimated decision rule by CODA-HO are very close to that under the true rule $d^{opt}$, with only one iteration. This supports the implementation technique discussed in Section \ref{sec:impl}.  


\subsection{Investigation with Multiple Intermediate Outcomes}\label{simu_multi_mid}

 We next consider $r=10$ and $s=2$ with the following three scenarios. 

\noindent \textbf{Scenario 3}: 
		$~~U^M(X)=
   			[{X^{(1)}}+2X^{(2)}, 0]^\top,
		~~C^M(X)=
   			[X^{(1)} X^{(2)}, 0]^\top,~~\\
		U^Y(X)=2{X^{(1)}}+X^{(2)},~~
		C^Y(X)=2X^{(1)} X^{(2)}.$

\noindent \textbf{Scenario 4}: 
		$~U^M(X)=
   			[0.5\{X^{(1)}\}^2+2X^{(2)}, 0]^\top, ~~
		C^M(X)=
   			[X^{(1)}  X^{(2)}, 0]^\top,\\
		U^Y(X)=2{X^{(1)}}+X^{(2)}, ~~
		C^Y(X)=2X^{(1)}  X^{(2)}.$

\noindent \textbf{Scenario 5}: 
		$~U^M(X)=
   			[{X^{(1)}}+2X^{(2)},
   			0.5\{X^{(1)}\}^2+2X^{(2)}]^\top,
		C^M(X)=
   			[X^{(1)} X^{(2)}, X^{(1)} X^{(2)}]^\top,\\
		U^Y(X)=2\cos\{X^{(1)}\}+X^{(2)}, ~~
		C^Y(X)=2 X^{(1)} X^{(2)}.$

 The true ODR for Scenarios 3 to 5 is the same as  $d^{opt}(X) = \mathbb{I}\{X^{(1)}{X^{(2)}}>0\}$, with the true value $V(d^{opt})$ as 0.999 for Scenarios 3 and 4 while 1.909 for Scenario 5, based on Monte Carlo approximations. Using a similar procedure introduced in Section \ref{simu_eval}, we apply CODA-HO, in comparison to `single baseline'. 
 The empirical results are summarized in Table \ref{table:4} for Scenarios 3 to 5, aggregated over 500 replications. 
%
{It can be seen from Table \ref{table:4} that CODA-HO performs reasonably better than the baseline procedure in terms of smaller variance under all scenarios. Specifically, in Scenario 3 with the baseline function linear in $X$, CODA-HO achieves a standard deviation of 0.095, against the larger standard deviation of 0.128 under `single baseline', with improved efficiency as 25.8\%, under $N_P=1000$. In Scenarios 4 and 5 with more complex non-linear baseline functions, CODA-HO outperforms `single baseline' by reducing the standard deviation as 16.4\% and 19.8\%, respectively, under $N_P=1000$. In addition, 
the estimated value function under the estimated ODR obtained by CODA-HO achieves better coverage probabilities in comparison to the corresponding estimators obtained using the primary sample solely under all scenarios when the sample size is small, $N_P = 500$, indicating a stronger capacity of the proposed method in handling many covariates by incorporating more samples with multiple mediators.}

  \begin{table}[!t]
\caption{Empirical results of the proposed CODA-HO method in comparison to the ODR based on the primary sample solely  under Scenarios 3 to 5.}\label{table:4} 
\begin{center}
\begin{tabular}{ccc|cc||cc|cc}
\hline\hline
		Method (Rule)&\multicolumn{2}{c|}{CODA ($d^{opt}$)} &\multicolumn{2}{c||}{CODA ($\widehat{d}$)} &\multicolumn{2}{c|}{ODR ($d^{opt}$)} &\multicolumn{2}{c}{ODR ($\widehat{d}_P$)} \\
		\hline
		$N_P=$& $500$ & $1000$& $ 500$ & $ 1000$& $ 500$ & $ 1000$& $ 500$ & $ 1000$\\
\hline\hline
		 \textbf{Scenario 3} $V(d^{opt})$ = & 0.999\\
		 \hline
Estimated $\widehat{V}(\cdot)$&0.983   & 0.986   & 1.037  & 1.021   & 0.984   & 0.980   & 1.072   & 1.038    \\
\hline
 $SD\{\widehat{V}(\cdot)\}$&0.130   & 0.093   & 0.128  & 0.093   & 0.182   & 0.126   & 0.180   & 0.125    \\\hline
$\Mean\{\widehat{\sigma}\}$&0.130   & 0.096   & 0.129   & 0.095  & 0.182   & 0.128   & 0.181   & 0.128    \\\hline
Coverage Probabilities &94.6\% & 95.8\% & 95.2\% & 94.8\% & 94.6\% & 94.6\% & 92.2\% & 94.8\%  \\
\hline
Improved Efficiency & 28.6\%  & 25.0\%  & 28.7\%  & 25.8\%   & / & /   & /   &/  \\
\hline\hline
		 \textbf{Scenario 4 } $V(d^{opt})$ =&  0.999\\
		 \hline
Estimated $\widehat{V}(\cdot)$&0.980   & 0.984   & 1.037   & 1.022   & 0.984   & 0.980   & 1.072   & 1.038    \\
\hline
 $SD\{\widehat{V}(\cdot)\}$&0.148   & 0.104   & 0.145   & 0.103   & 0.182   & 0.126   & 0.180   & 0.125    \\\hline
$\Mean\{\widehat{\sigma}\}$&0.148   & 0.107   & 0.148   & 0.107   & 0.182   & 0.128   & 0.181   & 0.128    \\\hline
Coverage Probabilities &95.0\% & 96.6\% & 95.8\% & 95.4\% & 94.6\% & 94.6\% & 92.2\% & 94.8\%  \\
\hline
Improved Efficiency & 18.7\%  & 16.4\%  & 18.2\%  & 16.4\%   & / & /   & /   &/  \\
\hline\hline
		 \textbf{Scenario 5}  $V(d^{opt})$ =& 1.909\\
		 \hline
Estimated $\widehat{V}(\cdot)$&1.898   & 1.895  & 1.977   & 1.948   & 1.898   & 1.889  & 2.004  & 1.962  \\
\hline
$SD\{\widehat{V}(\cdot)\}$&0.116   & 0.083   & 0.113   & 0.080   & 0.147   & 0.102   & 0.142  & 0.099    \\\hline
$\Mean\{\widehat{\sigma}\}$&0.116   & 0.084   & 0.116   & 0.085  & 0.150  & 0.106   & 0.150   & 0.106    \\\hline
Coverage Probabilities &94.8\% & 96.0\% & 92.2\% & 93.8\% & 95.2\% & 95.6\% & 91.0\% & 92.2\% \\\hline
 Improved Efficiency & 22.7\%  & 20.8\%  &  22.7\%  & 19.8\%   & / & /   & /   &/  \\
\hline 
\end{tabular}
\end{center}
\end{table}

 {\subsection{Evaluation on Calibrated Value Estimator for Heterogeneous Baseline Covariates}\label{simu_eval_hete}
 We next consider samples with heterogeneous baseline covariates generated by $ X_P^{(1)}, \cdots,X_P^{(r)}\\  {\sim} \text{Uniform}[-2,2],$ and $X_U^{(1)}, \cdots,X_U^{(r)} {\sim} \text{Uniform}[-1,1.5]$. All the rest settings are the same as Section \ref{simu_eval}. 
Using a similar procedure introduced in Section \ref{simu_eval}, we apply the proposed CODA-HE method ($\widehat{d}_R$), in comparison to 
 `single baseline' ($\widehat{d}_P$). The empirical results are summarized in Table \ref{table:hete2} 
 aggregated over 500 replications. The coverage probabilities are calculated based on the $95\%$  CI in \eqref{ciaipw_hete}.} 
 {It can be observed from Table \ref{table:hete2}  that CODA-HE is more efficient than  `single baseline', 
 in all scenarios, under heterogeneous baseline covariates. Specifically,  CODA-HE improves efficiency by 6.3\% in Scenario 1 and 8.8\% in Scenario 2 for $N_P = 1000$. In addition, the values under CODA-HE approach the true as the sample size $N_P$ increases, which yields $V (\widehat{d}) = 0.984$ in Scenario 1 ($V (d^{opt}) = 0.999$) and $V (\widehat{d}) = 1.238$ in Scenario 2 $(\max_{d\in \Pi_1}V(d) = 1.251)$ when $N_P = 1000$. Finally, the coverage probabilities of the value estimator obtained by CODA-HE are close to the nominal level, which support 
 the theoretical results in Theorem \ref{thm3_hete} for heterogeneous baseline covariates.} 

\begin{table}[!t]
\caption{Empirical results of the proposed CODA-HE method in comparison to the ODR based on the primary sample solely under Scenarios 1-2  with {heterogeneous} baseline covariates, where $\max_{d\in \Pi_1}V(d) = 1.251$ in Scenario 2.}\label{table:hete2}
\begin{center}
\begin{tabular}{ccc|cc||cc|cc}
\hline\hline
Method (Rule)	&\multicolumn{2}{c|}{CODA ($d^{opt}$)} &\multicolumn{2}{c||}{CODA ($\widehat{d}_R$)} &\multicolumn{2}{c|}{ODR ($d^{opt}$)} &\multicolumn{2}{c}{ODR ($\widehat{d}_P$)} \\
		\hline
	$N_P=$	& $500$ & $1000$& $ 500$ & $ 1000$& $ 500$ & $ 1000$& $ 500$ & $ 1000$\\
		\hline\hline
		 \textbf{Scenario 1}&\\
		  \hline
Plug-in Value $V(\cdot)$ &\multicolumn{2}{c|}{0.999}  & 0.975   & 0.984   & \multicolumn{2}{c|}{0.999}   & 0.967   & 0.977    \\
\hline
 		\hline
Estimated Value $\widehat{V}(\cdot)$&1.001  & 0.992   & 1.059   & 1.029    & 1.006  & 0.993  & 1.095   & 1.050   \\
\hline
 $SD\{\widehat{V}(\cdot)\}$&0.162   & 0.112  & 0.161   & 0.112   & 0.173   & 0.122   & 0.171   & 0.121    \\
\hline
 $\Mean\{\widehat{\sigma}\}$&0.172   &0.120  &0.171   &0.120  & 0.182   & 0.128   & 0.181   & 0.128    \\
\hline
Coverage Probabilities &96.8\% & 95.8\% & 97.0\% & 94.8\%  & 96.8\% & 96.0\% & 94.2\% & 94.2\%  \\
\hline 
\hline
 Improved Efficiency &5.5\%   & 6.3\%  &5.5\%   & 6.3\%& /      &  /     &    /   &  /      \\
 \hline
 $\widehat{\boldsymbol{\rho}}(\cdot)$&2.43  & 3.08  &2.39  & 3.06 & /      &  /     &    /   &  /      \\
\hline 
 $\widehat{\Sigma}_M(\cdot)$ &3.19 & 4.67   &3.17  & 4.65  &    /   &  /     &    /   &  /     \\
\hline\hline
\textbf{Scenario 2}	&\\
\hline 
Plug-in Value $V(\cdot)$ &\multicolumn{2}{c|}{1.333}  & 1.232   & 1.239  & \multicolumn{2}{c|}{1.333}    & 1.226   & 1.235   \\
\hline
 		\hline
Estimated Value $\widehat{V}(\cdot)$&1.325   & 1.320   & 1.317   & 1.288   & 1.329   & 1.322   & 1.350   & 1.312    \\
\hline
 $SD\{\widehat{V}(\cdot)\}$&0.140   & 0.103  & 0.139   & 0.102  & 0.156  & 0.118   & 0.154   & 0.116   \\
\hline
 $\Mean\{\widehat{\sigma}\}$&0.148   & 0.104   & 0.148   & 0.104   & 0.162   & 0.114  & 0.161   & 0.114   \\
\hline
Coverage Probabilities &96.0\% & 95.0\% & 95.8\% & 95.6\% & 95.4\% & 93.6\% & 94.2\% & 93.6\%  \\
\hline 
\hline
 Improved Efficiency &8.6\%   & 8.8\%  &8.1\%   & 8.8\%& /      &  /     &    /   &  /      \\
 \hline
 $\widehat{\boldsymbol{\rho}}(\cdot)$&2.72  & 3.44  & 2.68  & 3.39   &     /  &  /     &     /  & /       \\
\hline 
 $\widehat{\Sigma}_M(\cdot)$ &3.56  & 5.18   & 3.52 &5.12& /      &/       &     /  & /       \\
\hline 
\end{tabular}
\end{center}
\end{table}


\section{Real Data Analysis}\label{sec:5}

We illustrate the proposed method by application to data sources from the MIMIC-III 
and  the eICU data. 
{
We consider  $r=11$ common baseline covariates in both samples after dropping variables with high missing rates}: age (years), gender (0=female, 1=male), admission weights (kg), admission temperature (Celsius), Glasgow Coma Score (0-15), sodium amount (meq/L), glucose amount (mg/dL), blood urea nitrogen amount (mg/dL), creatinine amount (mg/dL), white blood cell count (E9/L), and total input amount (mL). Here, the treatment is coded as 1 if receiving the vasopressor, and 0 if receiving other medical supervisions such as IV fluid resuscitation. 
Intermediate outcomes include the total urine output (mL) and the cumulated balance (mL) of metabolism for both samples. The outcome of interest ($Y_P$) is 0 if a patient died due to sepsis and 1 if a patient is still alive, observed only in the primary sample. 
By deleting the abnormal values in two datasets, we form the primary sample of interest consisting of $N_P=10746$ subjects and the auxiliary sample of $N_U=7402$ subjects. 

\begin{figure}[!t] 
\centering
\begin{subfigure} 
  \centering
  \includegraphics[width=0.48\textwidth]{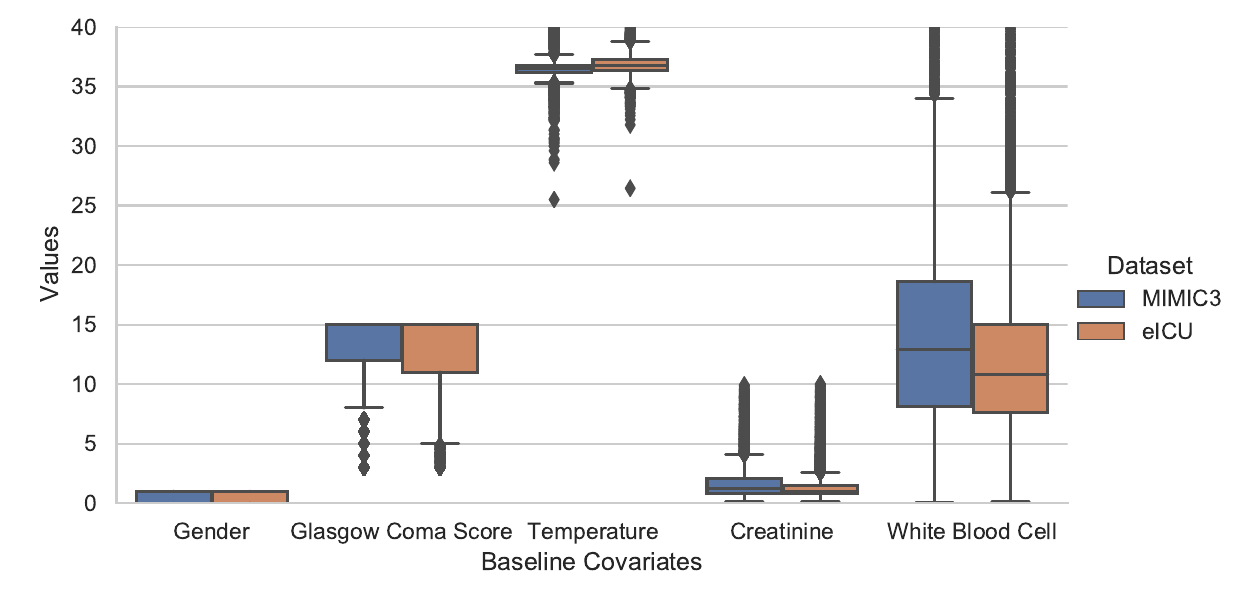}
\end{subfigure} 
\begin{subfigure} 
  \centering
  \includegraphics[width=.48\textwidth]{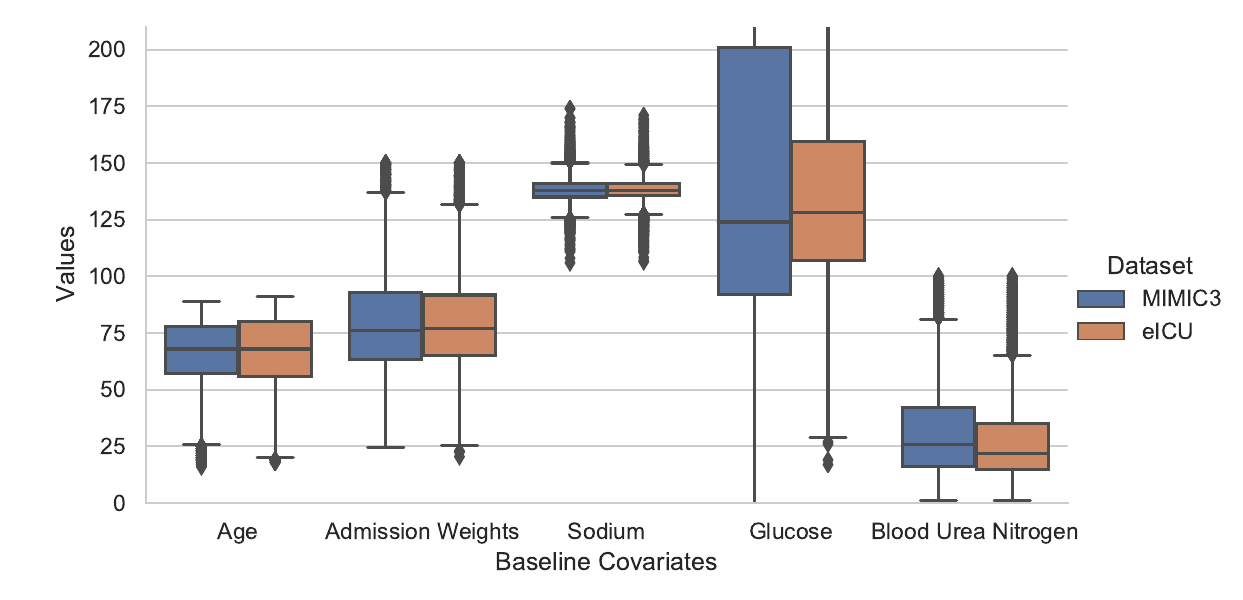}
\end{subfigure} 
  \caption{The box-plots for the shared baseline variables in the MIMIC-III data and the eICU data.}
  \label{fig:3}
\end{figure}

We illustrate the shared baseline variables in the MIMIC-III data and the eICU data in Fig. \ref{fig:3}. It can be seen from Fig. \ref{fig:3} that there exist multiple variables that have a distinct pattern in each sample, including Glasgow Coma Score, white blood cell count, glucose amount, and blood urea nitrogen amount. 
This shows some degree of heterogeneity in these two samples. To check the reasonability of (A4), we fit two intermediate outcomes on baseline covariates and the treatment in each sample based on the deep neural network. 
For each treatment-covariates pair $(x,a)$ in the support of the two samples, we predict $\Mean[M_P\mid X_P=x, A_P=a]$ based on the fitted deep neural network from the primary sample and $\Mean[M_U\mid X_U=x, A_U=a]$ based on the fitted deep neural network from the auxiliary sample. The Kullback-Leibler divergence between the fitted conditional means of two samples over the set of $(x,a)$ is 0.10 for the cumulated balance and 0.36 for the total output. 
This indicates the conditional mean estimators for intermediate outcomes in the two samples are close. 
Next, we apply CODA-HO and CODA-HE  on the two samples in comparison to  `single baseline', 
using a similar procedure introduced in Section \ref{simu_eval}. {We consider 
$N_P\in\{5000,10746\}$.} 
The results of the estimated value $\widehat{V}$, the estimated standard error $ \widehat{\sigma}$, the improved efficiency as the relative reduction in the estimated standard error of the CODA value estimator with respect to that of  `single baseline', 
and the number of the assignment to each treatment, are summarized in Table \ref{table:6}. In addition, we report the assignment matching rate between CODA and  `single baseline' 
in the last row of Table \ref{table:6}.

\begin{table}[!t]
\caption{The real data analysis under the proposed CODA method and the ODR method based on the primary sample solely.}\label{table:6}
\begin{center}
\begin{tabular}{cccc|ccc}
\hline\hline
Sample Size&\multicolumn{3}{c|}{ $N_P=5000$} &\multicolumn{3}{c}{ $N_P=10746$} \\
\hline		
Method & CODA-HO &  CODA-HE& ODR&  CODA-HO &  CODA-HE  & ODR \\
\hline
\hline
Estimated $\widehat{V}(\cdot)$   &0.204 &0.194&0.184&0.203 &0.200&0.192 \\
\hline
Estimated $\widehat{\sigma}$   &0.0090&0.0094& 0.0097 &0.0065&0.0067& 0.0068 \\
\hline
Improved Efficiency &7.2\%&3.1\%&/&4.4\%&1.5\%&/\\
\hline
Treatment 0&2967&2805&2671&5853&5665&5477\\
\hline
Treatment 1&2033&2195&2329&4893&5081&5269\\
\hline
Matching Rate & 87.5\%&93.7\%& /& 86.5\%&98.3\%& / \\
\hline
\end{tabular}
\end{center}
\end{table}

Based on Table \ref{table:6}, CODA performs reasonably better than 
`single baseline' 
under each different $N_P$. Specifically, with the full primary dataset ($N_P=10746$), {CODA-HO achieves a value of 0.203 with a smaller standard error of 0.0065, comparing to the value under the ODR as 0.192 with a standard error of 0.0068. 
The efficiency is improved by 4.4\% owing to CODA-HO. The rate of making the same decision between these two rules is 86.5\%. CODA-HO assigns 5853 patients to treatment 1 and 4893 patients to the control, which is consistent with the competitive nature of these two treatments. On the other hand, the proposed CODA-HE achieves a value of 0.200 with a slightly smaller standard error of 0.0067, in contrast to `single baseline'. 
The efficiency is improved by 1.5\% based on CODA-HE. The decision rule under CODA-HE is closer to the ODR. In addition, our CODAs could achieve a greater improvement in efficiency when the sample size is smaller, as 7.2\% for CODA-HO and 3.1\% for CODA-HE under $N_P=5000$. These findings are consistent with what we have observed in simulations, which demonstrate that CODA offers a statistically innovative and practically pragmatic tool for a more efficient optimal treatment decision making by integrating multiple data sources {from heterogeneous studies} with the limited outcome.

\section{Discussions and Extensions}\label{sec:6}

 {It is noted that the magnitude of efficiency gain of CODA depends on $\boldsymbol{\rho}(d)$ or $\boldsymbol{\rho}_R(d)$, the correlation between the value estimator of the outcome of interest and the value difference estimator of intermediate outcomes in the two samples. Specifically, for the case with homogeneous baseline covariates, we have
\begin{eqnarray}\nonumber
	\boldsymbol{\rho}(d) &=& 	E\Bigl({\mathbb{I}\{A_P=d(X_P)\}[Y_P - \mu_P\{X_P,d(X_P)\} ]\over{[A_P \pi_P (X_P)+(1-A_P)\{1-\pi_P(X_P)\}]^2}}   \times  [M_P - \theta\{X_P,d(X_P)\}]\Bigr)
	\\\label{theo_form_rho}
	& &+ E\Bigl( [\mu_P\{X_P,d(X_P)\} - V(d)]\times [\theta\{X_P,d(X_P)\} - W(d)]\Bigr), 
\end{eqnarray}	
while for the case with heterogeneous baseline covariates, we have
\begin{eqnarray*}
 \boldsymbol{\rho}_R(d) &=&E\Bigl({\mathbb{I}\{A_P=d(X_P)\}\over{[A_P \pi_P (X_P)+(1-A_P)\{1-\pi_P(X_P)\}]\times [A_P \pi (X_P)+(1-A_P)\{1-\pi(X_P)\}]}}   \\
	& & \times \frac{1}{r \{X_P,d(X_P),M_P\}} [Y_P - \mu_P\{X_P,d(X_P)\} ] \times  [M_P - \theta\{X_P,d(X_P)\}]\Bigr).
\end{eqnarray*}	
In general, $\boldsymbol{\rho}(d)$ tends to be larger than $\boldsymbol{\rho}_R(d)$ due to the second summation term in $\boldsymbol{\rho}(d)$ in \eqref{theo_form_rho}, which partly explains why CODA-HO has relatively larger efficiency gain than CODA-HE over the ODR obtained using the primary sample solely as observed in both simulations and the real data application.}



There are several possible extensions we may consider in future work. First, we only consider two treatment options in this paper, while in applications it is common to have more than two options for decision making. Thus, a more general method with multiple treatments or even continuous decisions is desirable. Second, we can extend the proposed CODA method to dynamic treatment decision making, where each subject successively receives a treatment followed by intermediate outcomes, however, the primary outcome of interest can be observed in the primary sample only. {Third, we only consider the setting where two samples share the same set of baseline covariates so that the ignorability assumption holds in both samples. In practice, different samples from heterogeneous studies may not have exactly the same set of baseline covariates. To be specific, let $X_P=[X_P^{(1)},\cdots, X_P^{(r_1)}]^\top$ denote $r_1$-dimensional individual's baseline covariates in the primary sample, and let $X_U=[X_U^{(1)},\cdots, X_U^{(r_2)}]^\top$ denote $r_2$-dimensional individual's baseline covariates in the auxiliary sample. The ignorability assumption holds for its own set of covariates in each sample. Suppose two samples share a same subset of baseline covariates with dimension $r_3 \leq \min(r_1, r_2)$, denoted as $X_C$. 
 Suppose that $X_C$ has the same joint distribution in two samples and the comparable
intermediate outcomes assumption holds for this common set of covariates $X_C$. 
 Then, we can modify the proposed calibrated value estimator 
by calibrating only with this common set of baseline covariates in two samples, but maintain searching the ODR based on whole available baseline covariates in the primary sample. We leave it for future research.}

    \newpage
\bibliographystyle{agsm}
\bibliography{mycite}

\newpage
\appendix
 \section{Technical Proofs}

In this appendix, we give technical proofs for the established lemmas and theorems. 
    
\subsection{Proof of Lemma \ref{lem_cio}}
The proof of Lemma \ref{lem_cio} consists of five steps as follows. We remark that the key ingredient of the proof lies in the law of iterated expectation together with assumptions (A1) and (A4) under the homogeneous baseline covariates, $X_P\sim X_U $.

(s1.) First, for $\Mean \{M_P^*(d)\}=\Mean [M_P^*(0) \{1-d(X_P)\}+M_P^*(1) d(X_P)]$, taking its iterated expectation on $\{X_P, A_P\}$, we have 
\begin{equation*}
\begin{split}
\Mean \{M_P^*(d)\}=&\Mean [M_P^*(0) \{1-d(X_P)\}+M_P^*(1) d(X_P)]\\
=&\Mean \Bigg[ \Mean \Bigg\{M_P^*(0) \{1-d(X_P)\}+M_P^*(1) d(X_P)\mid X_P=x,A_P=a\Bigg\}\Bigg]\\
=&\Mean \Bigg[ \Mean \Bigg\{M_P^*(0) \{1-d(x)\}+M_P^*(1) d(x)\mid X_P=x,A_P=a\Bigg\}\Bigg].
\end{split}
\end{equation*}

(s2.) By the assumption (A1) that $M_P= A_P M_P^{*}(1)  + (1-A_P)M_P^{*}(0)$, with the fact $A_P A_P=A_P$ and $A_P (1- A_P) = 0$ for $A_P \in \{0,1\}$, we have
\begin{equation*}
A_P M_P= A_P \{A_P M_P^{*}(1)  + (1-A_P)M_P^{*}(0)\} = M_P^{*}(1).
\end{equation*}
As such, we represent $M_P^{*}(0)$ as
\begin{equation*}
(1-A_P) M_P= (1-A_P) \{A_P M_P^{*}(1)  + (1-A_P)M_P^{*}(0)\} = M_P^{*}(0).
\end{equation*}

(s3.) By replacing $M_P^{*}(1)$ and $M_P^{*}(0)$ in the value function with results in (s2.), we have 
\begin{equation*}
\begin{split}
\Mean \{M_P^*(d)\}=&\Mean  \Bigg[ \Mean \Bigg\{(1-A_P) M_P \{1-d(x)\}+A_P M_P d(x)\mid X_P=x,A_P=a\Bigg\}\Bigg]\\
=&\Mean  \Bigg[ \Mean \Bigg\{(1-a) M_P \{1-d(x)\}+a M_P d(x)\mid X_P=x,A_P=a\Bigg\}\Bigg]\\
=&\Mean  \Bigg[  \Big[(1-a) \{1-d(x)\} +a d(x) \Big] \Mean \Big\{M_P\mid X_P=x,A_P=a\Big\} \Bigg].\\
\end{split}
\end{equation*}

(s4.) By the assumption (A4) that $\Mean[M_P\mid X_P=x, A_P=a]=\Mean[M_U\mid X_U=x,A_U=a]$, we have
\begin{equation*}
\begin{split}
\Mean \{M_P^*(d)\}=&\Mean \Bigg[  \Big[(1-a) \{1-d(x)\} +a d(x) \Big] \Mean \Big\{M_P\mid X_P=x,A_P=a\Big\} \Bigg]\\
=&\Mean  \Bigg[  \Big[(1-a) \{1-d(x)\} +a d(x) \Big] \Mean \Big\{M_U\mid X_U=x,A_U=a\Big\} \Bigg],
\end{split}
\end{equation*}
where the last step is valid since the expectation is taking over the same baseline distributions ($ {X}_P \sim  {X}_U$) on two sides. 

(s5.) Finally, the inverse procedure of steps (s3.) to (s1.) on the results in the step (s4.) leads to
\begin{equation*}
\Mean \{M_P^*(d)\}= \Mean [M_U^*(0) \{1-d(X_U)\}+M_U^*(1) d(X_U)] = \Mean \{M_U^*(d)\}.
\end{equation*}

The proof is hence completed.

\subsection{Proof of Lemma \ref{lem3}} 


We detail the proof for Lemma \ref{lem3} in this section. First, the term $\sqrt{N_P}  [\widehat{W}_{P}(d)  - \widehat{W}_{U}(d)  ] $ can be decomposed by
\begin{eqnarray}\label{decom_Me_Mu}
 \sqrt{N_P} \Big[\widehat{W}_{P}(d)  - \widehat{W}_{U}(d) \Big] =&&\sqrt{N_P} \Big[\widehat{W}_{P}(d)  - W(d)+  W(d)- \widehat{W}_{U}(d) \Big] \\\nonumber
=&& \sqrt{N_P} \Big[\widehat{W}_{P}(d)    - W(d) \Big] - \sqrt{t} \sqrt{N_U}  \Big[\widehat{W}_{U}(d) -W(d)\Big],
\end{eqnarray}
where $\sqrt{t} = {\sqrt{N_P  / N_U}}$ denotes the square root of the sample ratio. 

We next show the third line in \eqref{decom_Me_Mu} is asymptotically normal with mean zero. To do this, we define the independent and identically distributed double robust terms of intermediate outcomes in two samples as
\begin{equation*}
\boldsymbol{w}^{(i)}_P(d) \equiv    {\mathbb{I}\{d(X_{P,i})\} [M_{P,i} - \theta \{X_{P,i},d(X_{P,i})\} ]\over{A_{P,i} {\pi}_P (X_{P,i})+(1-A_{P,i})\{1-{\pi}_P(X_{P,i})\}}}   +  {\theta} \{X_{P,i},d(X_{P,i})\} ,
\end{equation*}
and  
\begin{equation*}
{\boldsymbol{w}}^{(i)}_U(d) \equiv   {\mathbb{I}\{d(X_{U,i})\} [M_{U,i}  - {\theta} \{X_{U,i},d(X_{U,i})\} ]\over{A_{U,i} {\pi}_U (X_{U,i})+(1-A_{U,i})\{1-{\pi}_U(X_{U,i})\}}}   + {\theta} \{X_{U,i},d(X_{U,i})\}  ,
\end{equation*}
where ${\pi}_P$ and ${\pi}_U$ are the true propensity score function in two samples, and ${\theta}$ is the true conditional mean function of the intermediate outcome given the covariates and the treatment.
 
Following the proof of Theorem 1 in the appendix B of \cite{rai2018statistical}, under assumptions (A6) and (A7. i, ii, and iii), given a decision rule $d(\cdot)$ that satisfies the assumption (A5), we have
\begin{equation}\label{pf_lem3_s1}
\widehat{W}_{P}(d) = {1\over N_P}\sum_{i=1}^{N_P} \boldsymbol{w}^{(i)}_P(d)  +o_p(N_P^{-1/2}),
\widehat{W}_{U}(d) = {1\over N_U}\sum_{i=1}^{N_U} \boldsymbol{w}^{(i)}_U(d)  +o_p(N_U^{-1/2}).
\end{equation}

Combining \eqref{pf_lem3_s1} with \eqref{decom_Me_Mu} yields that
\begin{eqnarray}\label{decom_Me_Mu_rep}
&&\sqrt{N_P} \Big[\widehat{W}_{P}(d)  - \widehat{W}_{U}(d) \Big] = \sqrt{N_P} \Big[\widehat{W}_{P}(d)    - W(d) \Big] - \sqrt{t} \sqrt{N_U}  \Big[\widehat{W}_{U}(d) -W(d)\Big],\\\nonumber
=&& \sqrt{N_P} \Big[{1\over N_P}\sum_{i=1}^{N_P} \boldsymbol{w}^{(i)}_P(d)    - W(d) \Big] - \sqrt{t} \sqrt{N_U}  \Big[{1\over N_U}\sum_{i=1}^{N_U} \boldsymbol{w}^{(i)}_U(d) -W(d)\Big]+o_p(1),
\end{eqnarray}
since $t \in(0, +\infty)$.

Based on the assumption (A4), $X_P \sim X_U$, and the central limit theorem, we have 
\begin{eqnarray}\label{normal_Me}
\sqrt{N_P} \Big[{1\over N_P}\sum_{i=1}^{N_P} \boldsymbol{w}^{(i)}_U(d)    - W(d) \Big] \rightsquigarrow N_s\Big\{\boldsymbol{0}_{s}, \Sigma_P(d)\Big\}, \\\label{normal_Mu}
\text{ and } \sqrt{N_U}  \Big[{1\over N_U}\sum_{i=1}^{N_U} \boldsymbol{w}^{(i)}_U(d) -W(d)\Big]  \rightsquigarrow N_s\Big\{\boldsymbol{0}_{s}, \Sigma_U(d)\Big\},
\end{eqnarray}
where $\boldsymbol{0}_{s}$ is $s$-dimensional zero vector, $\Sigma_P$ and $\Sigma_U$ are $s\times s$ matrices, and $ N_s(\cdot,\cdot)$ is the $s$-dimensional multivariate normal distribution.

Notice that the two samples ($P$ and $U$) are independently collected from two separate studies, and thus the independent and identically distributed double robust terms $\{\boldsymbol{w}^{(i)}_P(d) \}_{1\leq i \leq N_P}$ are independent of $\{\boldsymbol{w}^{(i)}_U(d) \}_{1\leq i \leq N_U}$. Hence, by noting the fact that the linear combination of two independent random variables having a normal distribution also has a normal distribution, based on \eqref{decom_Me_Mu_rep}, \eqref{normal_Me}, and \eqref{normal_Mu}, by Slutsky's theorem, with $T=\text{lim}_{N_P\to +\infty} t < +\infty$, we have 
 \begin{eqnarray}\label{normal_M_diff}
\sqrt{N_P} \Big[\widehat{W}_{P}(d)  - \widehat{W}_{U}(d) \Big]  \rightsquigarrow  N_s\Big\{\boldsymbol{0}_{s}, \Sigma_M(d)\Big\},
\end{eqnarray}
where $\Sigma_M(d) = \Sigma_P(d)+ T\Sigma_U(d)$ is a $s\times s$ matrix. The proof is hence completed.

\subsection{Proof of Lemma \ref{lem2_hete}} 


We detail the proof for Lemma \ref{lem2_hete} in this section. We focus on proving  the asymptotic normality of $\sqrt{n}  \widehat{W}_{1}(d) $, and the results for $\sqrt{n}  \widehat{W}_{0}(d)  $ can be shown in a similar manner. Let   \begin{eqnarray*}
\widehat{Z}_i = {R_i\over\widehat{r}\{X_i,d(X_i),M_i\} }{\mathbb{I}\{A_{i}=d(X_{i})\}\over{A_{i} \widehat{\pi}(X_{i})+(1-A_{i})\{1-\widehat{\pi}(X_{i})\}}}  [M_{i}  - \widehat{\theta} \{X_{i},d(X_{i})\} ]+ \widehat{\theta} \{X_{i},d(X_{i})\},
\end{eqnarray*} 
for $i=1,\cdots, n $, and thus $\widehat{W}_{1}(d)=   {1 / n}\sum_{i=1}^{n} \widehat{Z}_i $. Denote its counterpart as
\begin{eqnarray*}
Z_i = {R_i\over  {r}\{X_i,d(X_i),M_i\} }{\mathbb{I}\{A_{i}=d(X_{i})\}\over{A_{i}  {\pi}(X_{i})+(1-A_{i})\{1- {\pi}(X_{i})\}}}  [M_{i}  -  {\theta} \{X_{i},d(X_{i})\} ]+ {\theta} \{X_{i},d(X_{i})\}.
\end{eqnarray*} 
It is immediate from the central limit theorem that
\begin{eqnarray}\label{pf_lem2_hete_s1}
{1\over \sqrt{n}}  \sum_{i=1}^{n} {Z}_i   \rightsquigarrow N_s\Big\{W^*(d), \Sigma_{1}(d)\Big\},
\end{eqnarray} 
where $\Sigma_{1}$ is a $s\times s$ matrice presenting the asymptotic covariance matrice, and 
\begin{equation*}
W^*(d) =\Mean(Z_i)= \int E\{M\mid d(X),X\}\{P(R=1) f(E,X) + P(R=0) f(U,X)\} dX.
\end{equation*} 
According to \eqref{pf_lem2_hete_s1}, to show 
\begin{eqnarray}\label{pf_lem2_hete_s2}
\sqrt{n} \Big\{\widehat{W}_{1}(d)  -  W^*(d) \Big\} \rightsquigarrow N_s\Big\{\boldsymbol{0}_{s}, \Sigma_{1}(d)\Big\},
\end{eqnarray} 
it is sufficient to show
\begin{eqnarray}\label{pf_lem2_hete_s3}
{1\over \sqrt{n}}  \sum_{i=1}^{n} (\widehat{Z}_i - {Z}_i ) =o_p(1).
\end{eqnarray} 
 To this end, we define a middle term to assist our derivation as
\begin{eqnarray*}
\tilde{Z}_i = {R_i\over  {r}\{X_i,d(X_i),M_i\} }{\mathbb{I}\{A_{i}=d(X_{i})\}\over{A_{i}  \widehat{\pi}(X_{i})+(1-A_{i})\{1- \widehat{\pi}(X_{i})\}}}  [M_{i}  -  \widehat{\theta} \{X_{i},d(X_{i})\} ]+ \widehat{\theta} \{X_{i},d(X_{i})\}.
\end{eqnarray*} 
The rest of this section is focusing on proving \eqref{pf_lem2_hete_s3}, by the following two steps
\begin{eqnarray}\label{pf_lem2_hete_s4}
{1\over \sqrt{n}}  \sum_{i=1}^{n} (\widehat{Z}_i - \tilde{Z}_i )=o_p(1),
\end{eqnarray} 
and 
\begin{eqnarray}\label{pf_lem2_hete_s5}
{1\over \sqrt{n}}  \sum_{i=1}^{n} (\tilde{Z}_i - {Z}_i) =o_p(1).
\end{eqnarray} 
Combining \eqref{pf_lem2_hete_s4} and \eqref{pf_lem2_hete_s5} yields \eqref{pf_lem2_hete_s3}, and thus \eqref{pf_lem2_hete_s2} is proved. 

We focus on proving \eqref{pf_lem2_hete_s4} first. Since 
\begin{eqnarray}\nonumber 
\widehat{Z}_i - \tilde{Z}_i   &&=\left\{{R_i\over\widehat{r}\{X_i,d(X_i),M_i\}} -{R_i\over {r}\{X_i,d(X_i),M_i\}}\right\}{\mathbb{I}\{A_{i}=d(X_{i})\} [M_{i}  - \widehat{\theta} \{X_{i},d(X_{i})\} ]  \over{A_{i} \widehat{\pi}(X_{i})+(1-A_{i})\{1-\widehat{\pi}(X_{i})\}}} \\ \label{pf_lem2_hete_s6}
&&=\left\{{R_i(r_i  - \widehat{r}_i)\over\widehat{r}_i r_i }\right\}{\mathbb{I}\{A_{i}=d(X_{i})\} [M_{i}  - \widehat{\theta} \{X_{i},d(X_{i})\} ]  \over{A_{i} \widehat{\pi}(X_{i})+(1-A_{i})\{1-\widehat{\pi}(X_{i})\}}}  , 
\end{eqnarray} 
where ${r}_i\equiv{r}\{X_i,d(X_i),M_i\} $ and $ \widehat{r}_i\equiv \widehat{r}\{X_i,d(X_i),M_i\}$. We can further decompose \eqref{pf_lem2_hete_s6} by
\begin{eqnarray}\nonumber 
\widehat{Z}_i - \tilde{Z}_i   =&&\left\{{R_i(r_i  - \widehat{r}_i)\over\widehat{r}_i r_i }\right\}{\mathbb{I}\{A_{i}=d(X_{i})\} [M_{i} - {\theta} \{X_{i},d(X_{i})\}+ {\theta} \{X_{i},d(X_{i})\} - \widehat{\theta} \{X_{i},d(X_{i})\} ]  \over{A_{i} \widehat{\pi}(X_{i})+(1-A_{i})\{1-\widehat{\pi}(X_{i})\}}}   \\ \nonumber
 =&&\underbrace{[M_{i} - {\theta} \{X_{i},d(X_{i})\}  ]   {R_i(r_i  - \widehat{r}_i)\mathbb{I}\{A_{i}=d(X_{i})\}  \over{\widehat{r}_i r_i[A_{i} \widehat{\pi}(X_{i})+(1-A_{i})\{1-\widehat{\pi}(X_{i})\}]}} }_{\omega_{1,i}}\\ \nonumber
&&+\underbrace{(r_i  - \widehat{r}_i) [  {\theta} \{X_{i},d(X_{i})\} - \widehat{\theta} \{X_{i},d(X_{i})\} ]    {R_i\mathbb{I}\{A_{i}=d(X_{i})\}  \over{\widehat{r}_i r_i  [A_{i} \widehat{\pi}(X_{i})+(1-A_{i})\{1-\widehat{\pi}(X_{i})\}]}}}_{\omega_{2,i}}.
\end{eqnarray} 
We next bound ${\omega_{1,i}}$ and ${\omega_{2,i}}$ using empirical process, respectively. Here, by definition, we have $ {\theta} \{X_{i},d(X_{i})\} = \Mean \{M_i\mid X_{i},d(X_{i})\}$, thus,  
\begin{eqnarray}\label{pf_lem2_hete_s7}
\sqrt{n} \Mean_n \omega_{1,i}\leq \sqrt{n} \Mean_n  C_1 \left[M_i - {\theta} \{X_{i},d(X_{i})\}\right]=o_p(1),
\end{eqnarray}
where  $C_1$ is the bound of 
\begin{eqnarray*}
{R_i(r_i  - \widehat{r}_i)\mathbb{I}\{A_{i}=d(X_{i})\}  \over{\widehat{r}_i r_i[A_{i} \widehat{\pi}(X_{i})+(1-A_{i})\{1-\widehat{\pi}(X_{i})\}]}},
\end{eqnarray*}
under the positivity assumption.

Also, by assumptions (A7. v), we have
\begin{equation}\label{pf_lem2_hete_s8}
\sqrt{n} \Mean_n \omega_{2,i}\leq \sqrt{n} \Mean_n  C_2 (r_i  - \widehat{r}_i) \left[  {\theta} \{X_{i},d(X_{i})\} - \widehat{\theta} \{X_{i},d(X_{i})\} \right]=o_p(1),
\end{equation}
where  $C_2$ is the bound of 
\begin{eqnarray*}
{R_i\mathbb{I}\{A_{i}=d(X_{i})\}  \over{\widehat{r}_i r_i  [A_{i} \widehat{\pi}(X_{i})+(1-A_{i})\{1-\widehat{\pi}(X_{i})\}]}}.
\end{eqnarray*} 
Combining \eqref{pf_lem2_hete_s7} with \eqref{pf_lem2_hete_s8} yields  \eqref{pf_lem2_hete_s4}.

 Next, we show \eqref{pf_lem2_hete_s5}. Since 
\begin{eqnarray} \label{pf_lem2_hete_s9}
\tilde{Z}_i - {Z}_i   = {R_i\over r_i} && \left\{ \underbrace{{\mathbb{I}\{A_{i}=d(X_{i})\} [M_{i}  -  {\theta} \{X_{i},d(X_{i})\} ]  \over{A_{i}  {\pi}(X_{i})+(1-A_{i})\{1- {\pi}(X_{i})\}}} + {\theta} \{X_{i},d(X_{i})}_{\omega_{3,i}}\}\right.\\\nonumber 
&&  -\underbrace{ {\mathbb{I}\{A_{i}=d(X_{i})\} [M_{i}  -  \widehat{\theta} \{X_{i},d(X_{i})\} ]  \over{A_{i} \widehat{\pi}(X_{i})+(1-A_{i})\{1-\widehat{\pi}(X_{i})\}}} - \widehat{\theta} \{X_{i},d(X_{i})\}}_{\omega_{4,i}} \\\nonumber 
&&\left. + \widehat{\theta} \{X_{i},d(X_{i})\} -  {\theta} \{X_{i},d(X_{i})\}\right\}  +{\theta} \{X_{i},d(X_{i})\} -  \widehat{\theta} \{X_{i},d(X_{i})\}\\\nonumber 
= {R_i\over r_i}&&\left\{({\omega_{3,i}} - {\omega_{4,i}}) +    \widehat{\theta} \{X_{i},d(X_{i})\} -  {\theta} \{X_{i},d(X_{i})\}\right\}  +{\theta} \{X_{i},d(X_{i})\} -  \widehat{\theta} \{X_{i},d(X_{i})\} \\\nonumber
= {R_i\over r_i} &&({\omega_{3,i}} - {\omega_{4,i}}) +   \underbrace{({R_i\over r_i}-1) [\widehat{\theta} \{X_{i},d(X_{i})\} -  {\theta} \{X_{i},d(X_{i})\} ]}_{\omega_{5,i}},
\end{eqnarray} 
where ${\omega_{3,i}} - {\omega_{4,i}}$ is the regular doubly robust estimator \citep{zhang2012robust} for $i$-th subject minus its counterpart. Using the similar arguments in proving \eqref{pf_lem2_hete_s7}, under assumption (A7. iv), we can show that 
\begin{eqnarray}\label{pf_lem2_hete_s10}
\sqrt{n} \Mean_n {R_i\over r_i} ( \omega_{3,i}  - {\omega_{4,i}} ) =o_p(1).
\end{eqnarray}
On the other hand, since $r_i = P(R_i=1\mid X_i,d(X_i),M_i)$, we have 
\begin{eqnarray}\label{pf_lem2_hete_s11}
\sqrt{n} \Mean_n {\omega_{5,i}} \leq  \sqrt{n} \Mean_n  C_3 ({R_i\over r_i  } -1) =o_p(1),
\end{eqnarray}
where $C_3$ is the bound of $\widehat{\theta} \{X_{i},d(X_{i})\} -  {\theta} \{X_{i},d(X_{i})\} $. This together with \eqref{pf_lem2_hete_s10} yields \eqref{pf_lem2_hete_s5}. Thus, we complete the proof of \eqref{pf_lem2_hete_s2}.

\subsection{Proof of Lemma \ref{lem3_hete}} 


The proof for Lemma \ref{lem3_hete} is a direct result from Lemma \ref{lem2_hete}. Recall we have
\begin{equation*}
\begin{split}
\sqrt{n} \Big\{\widehat{W}_{1}(d)  -  W^*(d) \Big\} \rightsquigarrow N_s\Big\{\boldsymbol{0}_{s}, \Sigma_{1}(d)\Big\},   \text{ and }
\sqrt{n} \Big\{\widehat{W}_{0}(d)  -  W^*(d) \Big\} \rightsquigarrow N_s\Big\{\boldsymbol{0}_{s}, \Sigma_{0}(d)\Big\}.
\end{split}
\end{equation*}
Based on the property of normal distribution, we have
\begin{equation}\label{pf_lemma4_res}
\sqrt{n} \{\widehat{W}_{1}(d) - \widehat{W}_{0}(d)\}   \rightsquigarrow  N_s\Big\{\boldsymbol{0}_{s}, \Sigma_{R}(d)\Big\},
\end{equation}
where $\Sigma_{R}(d) $ a $s\times s$ asymptotic covariance matrix for term $\boldsymbol{\psi}_1^{(i)}(d) - \boldsymbol{\psi}_0^{(i)}(d)$, where
\begin{equation*}
\boldsymbol{\psi}_1^{(i)}(d) ={R_i\over{r}\{X_i,d(X_i),M_i\} }{\mathbb{I}\{A_{i}=d(X_{i})\}\over{A_{i} {\pi}(X_{i})+(1-A_{i})\{1-{\pi}(X_{i})\}}}  [M_{i}  - {\theta} \{X_{i},d(X_{i})\} ],
\end{equation*}
and
\begin{equation*}
\boldsymbol{\psi}_0^{(i)}(d) = {(1-R_i)\over1- {r}\{X_i,d(X_i),M_i\} }{\mathbb{I}\{A_{i}=d(X_{i})\}\over{A_{i} {\pi}(X_{i})+(1-A_{i})\{1-{\pi}(X_{i})\}}}  [M_{i}  - {\theta} \{X_{i},d(X_{i})\} ].
\end{equation*}
Here, $\boldsymbol{\psi}_1^{(i)}(d)$ is non-zero if $R=1$ (from the $P$ sample), and $\boldsymbol{\psi}_0^{(i)}(d)$ is non-zero if $R=0$ (from the $U$ sample). Noticing that the two samples ($P$ and $U$) are independently collected from two separate studies, we have 
\begin{equation*}
\Sigma_{R}(d) = \Sigma_{1P}(d) + \Sigma_{0U}(d),
\end{equation*}
where $\Sigma_{1P}(d)= \Var\{\boldsymbol{\psi}_1^{(i)}(d) \}$ and $\Sigma_{0U}(d)= \Var\{\boldsymbol{\psi}_0^{(i)}(d) \}$ are $s\times s$ asymptotic covariance matrices. The proof is hence completed.
    
\subsection{Proof of Theorem \ref{thm2}} 
The proof of Theorem \ref{thm2} consists of seven parts. We show the consistency of each proposed estimator in each part.

\textbf{Proof of results (i):}
 
 We first show the theoretical form of ${\sigma}_Y^2 (d) $ and then prove the consistency of $\widehat{\sigma}_Y^2 (d)$ to ${\sigma}_Y^2 (d) $.
Define the independent and identically distributed double robust term of the primary outcome of interest in the primary sample as
\begin{equation*}
 {v}^{(i)}_P(d) \equiv   {\mathbb{I}\{d(X_{P,i})\} [Y_{P,i} - {\mu}_P\{X_{P,i},d(X_{P,i})\} ] \over{A_{P,i}  {\pi}_P (X_{P,i})+(1-A_{P,i})\{1- {\pi}_P(X_{P,i})\}}}   +  {\mu}_P\{X_{P,i},d(X_{P,i})\}, 
\end{equation*}
where ${\pi}_P$ and ${\mu}_P$ are the true propensity score function and the true conditional mean function of the primary outcome of interest in the primary sample, respectively.

Similarly, following the proof of Theorem 1 in the appendix B of \cite{rai2018statistical}, under assumptions (A6) and (A7), given a decision rule $d(\cdot)$ that satisfies the assumption (A5), we have
\begin{equation}\label{pf_thm2_p1_s1}
\widehat{V}_{P}(d) = {1\over N_P}\sum_{i=1}^{N_P} {v}^{(i)}_P(d)  +o_p(N_P^{-1/2}).
\end{equation}

Using similar arguments in proving \eqref{decom_Me_Mu_rep}
 and \eqref{normal_Me}, by the central limit theorem and Slutsky's theorem, we can show that
 \begin{eqnarray}\label{decom_Y_rep}
 \sqrt{N_P} \Big[\widehat{V}_{P}(d)  - V(d) \Big] =\sqrt{N_P} \Big[{1\over N_P}\sum_{i=1}^{N_P} {v}^{(i)}_P(d)   - V(d) \Big] +o_p(1)\rightsquigarrow  N\Big\{0, \sigma_Y^2(d)\Big\},
\end{eqnarray} 
where $\sigma_Y^2(d) = \Mean[\{{v}_P(d)-V(d)\}^2]$ and 
 \begin{eqnarray*}
 {v}_P(d)={\mathbb{I}\{d(X_{P})\} [Y_{P} - {\mu}_P\{X_{P},d(X_{P})\} ] \over{A_{P}  {\pi}_P (X_{P})+(1-A_{P})\{1- {\pi}_P(X_{P})\}}}   +  {\mu}_P\{X_{P},d(X_{P})\}.
 \end{eqnarray*}

Thus, by the weak law of large number, it is immediate that
 \begin{eqnarray}\label{pf_thm2_p1_s2}
 {1\over N_P}\sum_{i=1}^{N_P}\{{v}_P^{(i)}(d) -V(d)\}^2 \overset{p}{\longrightarrow} \sigma_Y^2(d).
\end{eqnarray} 

We next decompose the proposed variance estimator by 
\begin{eqnarray}\label{pf_thm2_p1_s3}
\widehat{\sigma}_Y^2 (d)=&& {1\over N_P}\sum_{i=1}^{N_P}\{\widehat{v}_P^{(i)}(d) - \widehat{V}_P(d)\}^2\\\nonumber
   =&&{1\over N_P}\sum_{i=1}^{N_P}\{\widehat{v}_P^{(i)}(d) - {v}_P^{(i)}(d) + {v}_P^{(i)}(d) -{V}(d)+{V}(d)- \widehat{V}_P(d)\}^2.
      \end{eqnarray} 
      
      According to \eqref{decom_Y_rep}, we have $ \widehat{V}_P(d) ={V}(d) + o_p(1)$. Due to the assumption (A6) that covariates and outcomes are bounded,  the above \eqref{pf_thm2_p1_s3} yields
\begin{eqnarray*}
\widehat{\sigma}_Y^2 (d)= &&{1\over N_P}\sum_{i=1}^{N_P}\{\widehat{v}_P^{(i)}(d) - {v}_P^{(i)}(d) + {v}_P^{(i)}(d) -{V}(d)+o_p(1)\}^2 \\\nonumber
=&&{1\over N_P}\sum_{i=1}^{N_P}\left[ \{{v}_P^{(i)}(d) -{V}(d) \}^2+ \underbrace{\{\widehat{v}_P^{(i)}(d) - {v}_P^{(i)}(d) \} \{\widehat{v}_P^{(i)}(d) + {v}_P^{(i)}(d) -2{V}(d) \}}_{\eta_0} \right] + o_p(1). 
      \end{eqnarray*}

By noticing the term $\{\widehat{v}_P^{(i)}(d) + {v}_P^{(i)}(d) -2{V}(d) \}$ is bounded owing to the assumption (A6), based on the results in \eqref{pf_thm2_p1_s1}, we have ${ N_P}^{-1}\sum_{i=1}^{N_P}  \eta_0 = o_p(1)$. This gives
\begin{eqnarray}\label{pf_thm2_p1_s4}
\widehat{\sigma}_Y^2 (d)= &&{1\over N_P}\sum_{i=1}^{N_P}\{\widehat{v}_P^{(i)}(d) - \widehat{V}_P(d)\}^2={1\over N_P}\sum_{i=1}^{N_P} \{{v}_P^{(i)}(d) -{V}(d) \}^2 + o_p(1).
      \end{eqnarray} 

Combining \eqref{pf_thm2_p1_s4} with \eqref{pf_thm2_p1_s2} yields $\widehat{\sigma}_Y^2 (d) ={\sigma}_Y^2 (d) +o_p(1).$ The proof of results (i) is hence completed.

\textbf{Proof of results (ii):}
 
 Similarly, we first show the theoretical form of $\boldsymbol{\rho}(d)$ and then prove the consistency of $\widehat{\boldsymbol{\rho}}(d) $ to $\boldsymbol{\rho}(d)$. Aware that the joint distribution of normal random variables is still normal, we have the following results based on Lemma \ref{lem1} and Lemma \ref{lem3}. 
 \begin{equation*}  
  \begin{split}
&\sqrt{N_P}		 \begin{bmatrix}
   			\widehat{V}_{P}(d)  -  V(d) \\
   			\widehat{W}_{P}(d)  - \widehat{W}_{U}(d) 
			\end{bmatrix}
	 \rightsquigarrow N_{s+1} \Bigg\{\boldsymbol{0}_{s+1}, \begin{bmatrix}
   			\sigma_Y^2(d) , \boldsymbol{\rho}(d)^\top\\
   			\boldsymbol{\rho}(d), \Sigma_M(d)
			\end{bmatrix}\Bigg\}, \quad \forall d(\cdot),
\end{split}
\end{equation*}
where $\boldsymbol{\rho}(d)$ is the $s\times 1$ asymptotic covariance vector between the value estimator of the outcome of interest in the primary sample and the differences of the value estimators of intermediate outcomes between two samples.

Recall the results in \eqref{decom_Me_Mu_rep} and \eqref{decom_Y_rep} that
 \begin{eqnarray}\label{pf_thm2_p2_s1}
  \sqrt{N_P} \Big[\widehat{V}_{P}(d)  - V(d) \Big] =&&\sqrt{N_P} \Big[{1\over N_P}\sum_{i=1}^{N_P} {v}^{(i)}_P(d)   - V(d) \Big]+o_p(1),\\\nonumber
 \sqrt{N_P} \Big[\widehat{W}_{P}(d)  - \widehat{W}_{U}(d) \Big] =&&\sqrt{N_P} \Big[{1\over N_P}\sum_{i=1}^{N_P} \boldsymbol{w}^{(i)}_P(d)    - W(d) \Big]\\\nonumber
&& - \sqrt{t} \sqrt{N_U}  \Big[{1\over N_U}\sum_{i=1}^{N_U} \boldsymbol{w}^{(i)}_U(d) -W(d)\Big]+ o_p(1).
\end{eqnarray}  

The asymptotic covariance between $\sqrt{N_P} \Big[\widehat{V}_{P}(d)  - V(d) \Big]$ and $\sqrt{N_P} \Big[\widehat{W}_{P}(d)  - \widehat{W}_{U}(d) \Big] $ comes from the correlation between $\{v^{(i)}_P(d) \}_{1\leq i \leq N_P}$ and $[\{\boldsymbol{w}^{(i)}_P(d) \}_{1\leq i \leq N_P}, \{\boldsymbol{w}^{(i)}_U(d) \}_{1\leq i \leq N_U}]$.

By noticing that the two samples ($P$ and $U$) are independently collected from two separate studies, the independent and identically distributed double robust terms $\{v^{(i)}_P(d) \}_{1\leq i \leq N_P}$ are independent of $\{\boldsymbol{w}^{(i)}_U(d) \}_{1\leq i \leq N_U}$.

Using similar arguments in proving \eqref{normal_M_diff} and \eqref{decom_Y_rep}, we have
 \begin{eqnarray}\label{rho_form}
\boldsymbol{\rho}(d)=\Mean\left [\Big\{{v}_P(d) - {V}(d) \Big\} \Big\{{\boldsymbol{w}} _P(d) -  W(d) \Big\}\right],
\end{eqnarray} 
where 
 \begin{eqnarray*}
\boldsymbol{w}_P(d) \equiv    {\mathbb{I}\{d(X_{P})\} [M_{P} - \theta \{X_{P},d(X_{P})\} ]\over{A_{P} {\pi}_P (X_{P})+(1-A_{P})\{1-{\pi}_P(X_{P})\}}}   +  {\theta} \{X_{P},d(X_{P})\} .
 \end{eqnarray*} 
Thus, by the weak law of large number, it is immediate that
 \begin{eqnarray}\label{pf_thm2_p2_s2}
 {1\over N_P}\sum_{i=1}^{N_P}\Big\{{v}_P^{(i)}(d) - {V}(d) \Big\} \Big\{{\boldsymbol{w}}^{(i)}_P(d) -  W(d) \Big\}  \overset{p}{\longrightarrow} \boldsymbol{\rho}(d).
\end{eqnarray} 

 Next, to show the consistency of $\widehat{\boldsymbol{\rho}}(d)$ to $\boldsymbol{\rho}(d)$, we decompose $\widehat{\boldsymbol{\rho}}(d)$ as
   \begin{eqnarray} \label{pf_thm2_p2_s3}
 \widehat{\boldsymbol{\rho}}(d)  =&& {1\over N_P}\sum_{i=1}^{N_P} \Big\{\widehat{v}_P^{(i)}(d) -\widehat{V}_P(d) \Big\} \Big\{\widehat{\boldsymbol{w}}^{(i)}_P(d) - \widehat{W}_P(d) \Big\}\\\nonumber
    =&&{1\over N_P}\sum_{i=1}^{N_P} \Big\{\widehat{v}_P^{(i)}(d) -{v}_P^{(i)}(d)+ {v}_P^{(i)}(d)- V(d) + V(d)- \widehat{V}_P(d) \Big\} \\\nonumber
    &&~~~~~~~~~~~~~\times \Big\{\widehat{\boldsymbol{w}}^{(i)}_P(d) - {\boldsymbol{w}}^{(i)}_P(d)+ {\boldsymbol{w}}^{(i)}_P(d)-W(d)+ W(d) - \widehat{W}_P(d) \Big\}.
\end{eqnarray} 

Using similar arguments in proving \eqref{pf_thm2_p1_s4}, we can show $$\widehat{\boldsymbol{\rho}}(d) = {1\over N_P}\sum_{i=1}^{N_P}\Big\{{v}_P^{(i)}(d) - {V}(d) \Big\} \Big\{{\boldsymbol{w}}^{(i)}_P(d) -  W(d) \Big\}+o_p(1).$$
 This together with \eqref{pf_thm2_p2_s2} proves results (ii). The proof is hence completed.

\textbf{Proof of results (iii):}

  We already established the theoretical form of $\Sigma_M = \Sigma_P(d)+ T\Sigma_U(d)$ in the proof of Lemma \ref{lem3}, where $T=\text{lim}_{N_P\to +\infty} t < +\infty$. 
  
  Following the similar arguments in establishing \eqref{pf_thm2_p1_s2}, we have
   \begin{eqnarray} \label{pf_thm2_p3_s1}
  &&  {1\over N_P}\sum_{i=1}^{N_P} \Big\{{\boldsymbol{w}}^{(i)}_P(d) -{W}(d) \Big\}{\Big\{{\boldsymbol{w}}^{(i)}_P(d) - {W}(d)\Big\}}^\top  \overset{p}{\longrightarrow} {\Sigma}_P(d),  \\\nonumber
  && {1\over N_U}\sum_{i=1}^{N_U} \Big\{{\boldsymbol{w}}^{(i)}_U(d) -{W}(d) \Big\}{\Big\{{\boldsymbol{w}}^{(i)}_U(d) - {W}(d)\Big\}}^\top \overset{p}{\longrightarrow} {\Sigma}_U(d).
\end{eqnarray} 
  
To show the consistency of $\widehat{\Sigma}_M(d)$ to $\Sigma_M$, we decompose $\widehat{\Sigma}_M(d)$ as
  \begin{eqnarray} 
  \widehat{\Sigma}_M(d) =&&{1\over N_P}\sum_{i=1}^{N_P} \Big\{\widehat{\boldsymbol{w}}^{(i)}_P(d) -{\boldsymbol{w}}^{(i)}_P(d)+ {\boldsymbol{w}}^{(i)}_P(d)- W(d)+W(d)-\widehat{W}_P(d) \Big\}\\\nonumber
 &&~~~~~~~~~~~~ \times\Big\{\widehat{\boldsymbol{w}}^{(i)}_P(d) -{\boldsymbol{w}}^{(i)}_P(d)+ {\boldsymbol{w}}^{(i)}_P(d)- W(d)+W(d)-\widehat{W}_P(d) \Big\}^\top \\\nonumber
  && + t {1\over N_U}\sum_{i=1}^{N_U} \Big\{\widehat{\boldsymbol{w}}^{(i)}_U(d) -{\boldsymbol{w}}^{(i)}_U(d)+{\boldsymbol{w}}^{(i)}_U(d) - W(d)+W(d) + \widehat{W}_U(d) \Big\}\\\nonumber
 &&~~~~~~~~~~~~ \times{\Big\{\widehat{\boldsymbol{w}}^{(i)}_U(d) -{\boldsymbol{w}}^{(i)}_U(d)+{\boldsymbol{w}}^{(i)}_U(d) - W(d)+W(d) + \widehat{W}_U(d) \Big\}}^\top.
\end{eqnarray} 

Using similar arguments in proving \eqref{pf_thm2_p1_s4}, we can show
  \begin{eqnarray*} 
  && \widehat{\Sigma}_M(d) ={1\over N_P}\sum_{i=1}^{N_P} \Big\{{\boldsymbol{w}}^{(i)}_P(d) -{W}_P(d) \Big\}{\Big\{{\boldsymbol{w}}^{(i)}_P(d) - {W}_P(d)\Big\}}^\top \\\nonumber
  &&~~~~~~~~~~~~~+ t {1\over N_U}\sum_{i=1}^{N_U} \Big\{{\boldsymbol{w}}^{(i)}_U(d) -{W}_U(d) \Big\}{\Big\{{\boldsymbol{w}}^{(i)}_U(d) - {W}_U(d)\Big\}}^\top+o_p(1).
  \end{eqnarray*} 
 This together with \eqref{pf_thm2_p3_s1} proves results (iii). The proof is hence completed.

\textbf{Proof of results (iv):}
 
The proof of this part can be shown using similar arguments as in the proof of results (ii). We first show the theoretical form of $\boldsymbol{\rho}_R(d)$ and then prove the consistency of $\widehat{\boldsymbol{\rho}}_R(d) $ to $\boldsymbol{\rho}_R(d)$. Recall the asymptotic joint normality that
 \begin{equation*}  
  \begin{split}
&\sqrt{N_P}		 \begin{bmatrix}
   			\widehat{V}_{P}(d)  -  V(d) \\
   			\sqrt{{n\over N_P}}\{\widehat{W}_{1}(d) - \widehat{W}_{0}(d)\} \end{bmatrix}
	 \rightsquigarrow N_{s+1} \Bigg\{\boldsymbol{0}_{s+1}, \begin{bmatrix}
   			\sigma_Y^2(d) , \boldsymbol{\rho}_{R}(d)^\top\\
   			\boldsymbol{\rho}_{R}(d), \Sigma_{R}(d)
			\end{bmatrix}\Bigg\},    \text{ for all } d(\cdot),
\end{split}
\end{equation*}
we have $\boldsymbol{\rho}_{R}(d)$ is the $s\times 1$ asymptotic correlation vector between the value estimator of the outcome of interest in the primary sample ($\sqrt{N_P}  [\widehat{V}_{P}(d)  - V(d)  ]$) and the new rebalanced value difference estimator of intermediate outcomes between two samples ($\sqrt{n}  [\widehat{W}_1(d)  - \widehat{W}_0(d)  ]$).

Recall the results in \eqref{decom_Me_Mu_rep} and \eqref{pf_lemma4_res}, we have
 \begin{eqnarray*}
  \sqrt{N_P} \Big[\widehat{V}_{P}(d)  - V(d) \Big] =&&\sqrt{N_P} \Big[{1\over N_P}\sum_{i=1}^{N_P} {v}^{(i)}_P(d)   - V(d) \Big]+o_p(1),\\\nonumber
  \sqrt{n} \Big[\widehat{W}_{1}(d) - \widehat{W}_{0}(d)\Big]   =&&\sqrt{n} \Big[{1\over n}\sum_{i=1}^{n}\{\boldsymbol{\psi}_1^{(i)}(d) -\boldsymbol{\psi}_0^{(i)}(d) \} \Big]+o_p(1),
\end{eqnarray*}  
where
\begin{equation*}
\boldsymbol{\psi}_1^{(i)}(d) ={R_i\over{r}\{X_i,d(X_i),M_i\} }{\mathbb{I}\{A_{i}=d(X_{i})\}\over{A_{i} {\pi}(X_{i})+(1-A_{i})\{1-{\pi}(X_{i})\}}}  [M_{i}  - {\theta} \{X_{i},d(X_{i})\} ],
\end{equation*}
and
\begin{equation*}
\boldsymbol{\psi}_0^{(i)}(d) = {(1-R_i)\over1- {r}\{X_i,d(X_i),M_i\} }{\mathbb{I}\{A_{i}=d(X_{i})\}\over{A_{i} {\pi}(X_{i})+(1-A_{i})\{1-{\pi}(X_{i})\}}}  [M_{i}  - {\theta} \{X_{i},d(X_{i})\} ].
\end{equation*}
The asymptotic covariance $\boldsymbol{\rho}_{R}(d)$ thus comes from the correlation between $\{v^{(i)}_P(d) \}_{1\leq i \leq N_P}$ and $[\{\boldsymbol{\psi}_1^{(i)}(d) \}_{1\leq i \leq n}, \{\boldsymbol{\psi}_0^{(i)}(d) \}_{1\leq i \leq n}]$. Here, $\boldsymbol{\psi}_1^{(i)}(d)$ is non-zero if $R=1$ (from the $P$ sample), and $\boldsymbol{\psi}_0^{(i)}(d)$ is non-zero if $R=0$ (from the $U$ sample). By noticing that the two samples ($P$ and $U$) are independently collected from two separate studies, the independent and identically distributed double robust terms $\{v^{(i)}_P(d) \}_{1\leq i \leq N_P}$ are independent of $[\{\boldsymbol{\psi}_1^{(i)}(d) \}_{(N_P+1)\leq i \leq n}, \{\boldsymbol{\psi}_0^{(i)}(d) \}_{1\leq i \leq n}]$.

Using similar arguments in proving \eqref{pf_thm2_p2_s2}, by the weak law of large number, it is immediate that
 \begin{eqnarray*}
 {1\over N_P}\sum_{i=1}^{N_P}\Big\{{v}_P^{(i)}(d) - {V}(d) \Big\}  \sqrt{N_P/n} \boldsymbol{\psi}_1^{(i)}(d)  \overset{p}{\longrightarrow} \boldsymbol{\rho}_R(d),
\end{eqnarray*} 
by noticing the term $\boldsymbol{\psi}_1^{(i)}(d) $ is mean zero.

Therefore, by decomposing $\widehat{\boldsymbol{\rho}}_R(d)$ in a similar manner as in \ref{pf_thm2_p2_s3}, we can show the consistency of $\widehat{\boldsymbol{\rho}}_R(d)$ to $\boldsymbol{\rho}_R(d)$. The proof for results (iv) is hence completed.

\textbf{Proof of results (v):}

This part can be easily shown using similar arguments as in the proof of results (iii). Recall the results in \eqref{pf_lemma4_res}, we have $\Sigma_{R}(d) $ is a $s\times s$ asymptotic covariance matrix for term $\boldsymbol{\psi}_1^{(i)}(d) - \boldsymbol{\psi}_0^{(i)}(d)$. Since $\boldsymbol{\psi}_1^{(i)}(d) - \boldsymbol{\psi}_0^{(i)}(d)$ is equivalent to  $\boldsymbol{w}_1^{(i)}(d) - \boldsymbol{w}_0^{(i)}(d)$ where
\begin{equation*}
\boldsymbol{w}_1^{(i)}(d) ={R_i\over{r}\{X_i,d(X_i),M_i\} }{\mathbb{I}\{A_{i}=d(X_{i})\}  [M_{i}  - {\theta} \{X_{i},d(X_{i})\} ]\over{A_{i} {\pi}(X_{i})+(1-A_{i})\{1-{\pi}(X_{i})\}}}   + {\theta} \{X_{i},d(X_{i})\}, 
\end{equation*}
and
\begin{equation*}
\boldsymbol{w}_0^{(i)}(d) = {(1-R_i)\over1- {r}\{X_i,d(X_i),M_i\} }{\mathbb{I}\{A_{i}=d(X_{i})\}  [M_{i}  - {\theta} \{X_{i},d(X_{i})\} ]\over{A_{i} {\pi}(X_{i})+(1-A_{i})\{1-{\pi}(X_{i})\}}}   + {\theta} \{X_{i},d(X_{i})\},
\end{equation*}
  we can show the theoretical form of $\Sigma_R$ following the similar arguments in establishing \eqref{pf_thm2_p1_s2} and \eqref{pf_thm2_p3_s1} as
    \begin{eqnarray}\label{pf_thm2_p5new_s1}
 {1\over n}\sum_{i=1}^{n} \Big\{{\boldsymbol{w}}^{(i)}_1(d) - \boldsymbol{w}^{(i)}_0(d)  \Big\}\Big\{\boldsymbol{w}^{(i)}_1(d) - \boldsymbol{w}^{(i)}_0(d) \Big\}^\top \overset{p}{\longrightarrow} {\Sigma}_R(d).
\end{eqnarray} 
Therefore, using similar arguments in proving \eqref{pf_thm2_p1_s4}, we can show
  \begin{eqnarray*} 
\widehat{\Sigma}_R(d) = {\Sigma}_R(d) +o_p(1).
  \end{eqnarray*} 
The proof of results (v) is hence completed.

\textbf{Proof of results (vi):}

We next show the consistency of the proposed calibrated value estimator $ \widehat{V}(d)$ to the true value $V(d) $. Recall \eqref{value_CODA} that
 \begin{equation*}
 \widehat{V}(d) = \widehat{V}_{P}(d) - \widehat{\boldsymbol{\rho}}(d)^\top \widehat{\Sigma}_M^{-1}(d) \{\widehat{W}_{P}(d)  - \widehat{W}_{U}(d)  \},
\end{equation*}
where $ \widehat{\boldsymbol{\rho}}(d) $ is the estimator for $\boldsymbol{\rho}(d)$, and $\widehat{\Sigma}_M(d)$ is the estimator for $\Sigma_M(d)$.
 
Based on the established results (ii) and (iii), with the assumption (A6) that covariates and outcomes are bounded, we have 
 \begin{equation*}
 \widehat{V}(d) =  \widehat{V}_{P}(d)- {\boldsymbol{\rho}}(d)^\top {\Sigma}_M^{-1}(d) \{\widehat{W}_{P}(d)  - \widehat{W}_{U}(d)  \} +o_p(1),
\end{equation*}

According to \eqref{decom_Y_rep} and \eqref{normal_M_diff}, we have $ \widehat{V}_P(d) ={V}(d) + o_p(1)$ and $\widehat{W}_{P}(d)  - \widehat{W}_{U}(d) = o_p(1)$ under $X_P \sim X_U$. Hence, it is immediate that  
 \begin{equation*}
 \widehat{V}(d) =  {V}(d) +o_p(1) = {V}(d)+o_p(1).
\end{equation*} 
The proof of results (vi) is hence completed.

\textbf{Proof of results (vii):}

Lastly, the consistency of the proposed calibrated value estimator $ \widehat{V}_R(d)$ to the true value $V(d) $ can be shown using the similar arguments in the proof of results (vi) by utilizing definition of  $ \widehat{V}_R(d)$, the results (iv) and (v), and \eqref{decom_Y_rep} and \eqref{pf_lemma4_res}. We omit the details for brevity.

\subsection{Proof of Theorem \ref{thm3}}\label{sec:pf_thm3}
In this section, we prove the asymptotic normality of the proposed calibrated value estimator. The proof consists of three parts. In part 1, we aim to show
 \begin{equation}\label{thm3_p1_mid1} 
\widehat{V}(\widehat{d}) =  {V}_n(\widehat{d}) +o_p(N_P^{-1/2}),
\end{equation} 
where 
 \begin{equation*}
 {V}_n(\widehat{d}) = {1\over N_P}\sum_{i=1}^{N_P} {v}_P^{(i)}(\widehat{d}) -  {\boldsymbol{\rho}}(\widehat{d})^\top {\Sigma}_M^{-1}(\widehat{d}) \left\{ {1\over N_P}\sum_{i=1}^{N_P} \boldsymbol{w}^{(i)}_P(\widehat{d})  - {1\over N_U}\sum_{i=1}^{N_U} \boldsymbol{w}^{(i)}_U(\widehat{d}) \right\}.
\end{equation*} 

Then, in the second part, we establish
 \begin{equation*}
 {V}_n(\widehat{d}) =  {V}_n(d^{opt}) +o_p(N_P^{-1/2}).
\end{equation*}

The above two steps yields that
 \begin{equation}\label{thm3_p1_mid2}
\widehat{V}(\widehat{d}) = {V}_n(d^{opt}) + o_p(N_P^{-1/2}).
\end{equation} 

Lastly, based on \eqref{thm3_p1_mid2}, we show
 \begin{equation*}  
\sqrt{N_P}\Big\{\widehat{V}(\widehat{d})-V(d^{opt})\Big\}   \rightsquigarrow N\Big[ 0, \sigma_Y^2(d^{opt}) - \boldsymbol{\rho}(d^{opt})^\top\Sigma_M^{-1}(d^{opt}) \boldsymbol{\rho}(d^{opt})  \Big]. 
\end{equation*}

\textbf{Proof of Part 1:}
Based on \eqref{pf_thm2_p1_s1}, we have $\widehat{V}_{P}(\widehat{d}) = { N_P^{-1}}\sum_{i=1}^{N_P} {v}^{(i)}_P(\widehat{d})  +o_p(N_P^{-1/2})$. Combining this with the definition of $\widehat{V}(\widehat{d})$ yields that 
 \begin{equation}\label{thm3_p1_s1} 
 \widehat{V}(\widehat{d}) ={1\over N_P}\sum_{i=1}^{N_P} {v}^{(i)}_P(\widehat{d})   -\underbrace{\widehat{\boldsymbol{\rho}}(\widehat{d})^\top \widehat{\Sigma}_M^{-1}(\widehat{d}) \{\widehat{W}_{P}(\widehat{d})  - \widehat{W}_{U}(\widehat{d})  \}}_{\eta_1} +o_p(N_P^{-1/2}),
\end{equation}

Thus, to show \eqref{thm3_p1_mid1}, it is sufficient to show
 \begin{equation}\label{thm3_p1_eta1} 
\eta_1=  {\boldsymbol{\rho}}(\widehat{d})^\top {\Sigma}_M^{-1}(\widehat{d}) \left\{ {1\over N_P}\sum_{i=1}^{N_P} \boldsymbol{w}^{(i)}_P(\widehat{d})  - {1\over N_U}\sum_{i=1}^{N_U} \boldsymbol{w}^{(i)}_U(\widehat{d}) \right\}+o_p(N_P^{-1/2}).
\end{equation}

According to  \eqref{normal_M_diff}, we have
 \begin{equation*} 
\widehat{W}_{P}(\widehat{d})  - \widehat{W}_{U}(\widehat{d})  = O_p(N_P^{-1/2}).
\end{equation*}

This together with results (ii) in Theorem \ref{thm2} that $\widehat{\boldsymbol{\rho}}(d) =\boldsymbol{\rho}(d)+o_p(1)$ yields
 \begin{eqnarray*}  
 \eta_1=&&\widehat{\boldsymbol{\rho}}(\widehat{d})^\top \widehat{\Sigma}_M^{-1}(\widehat{d}) \{\widehat{W}_{P}(\widehat{d})  - \widehat{W}_{U}(\widehat{d})  \} \\\nonumber
 =&&\{{\boldsymbol{\rho}}(\widehat{d})^\top +o_p(1)\} \widehat{\Sigma}_M^{-1}(\widehat{d})  \{\widehat{W}_{P}(\widehat{d})  - \widehat{W}_{U}(\widehat{d})  \}  \\\nonumber
  =&& {\boldsymbol{\rho}}(\widehat{d})^\top  \widehat{\Sigma}_M^{-1}(\widehat{d})    \{\widehat{W}_{P}(\widehat{d})  - \widehat{W}_{U}(\widehat{d})  \} +o_p(1) \widehat{\Sigma}_M^{-1}   (\widehat{d})   O_p(N_P^{-1/2}).
  \end{eqnarray*}
Since  $O_p(N_P^{-1/2})o_p(1) = o_p(N_P^{-1/2})$ and  $\widehat{\Sigma}_M^{-1}$ is bounded, we have the above equation as
   \begin{eqnarray}\label{thm3_p1_s2}  
 \eta_1= {\boldsymbol{\rho}}(\widehat{d})^\top  \widehat{\Sigma}_M^{-1}(\widehat{d})    \{\widehat{W}_{P}(\widehat{d})  - \widehat{W}_{U}(\widehat{d})  \}  + o_p(N_P^{-1/2}).
  \end{eqnarray}
  
 Similarly, combining results (iii) in Theorem \ref{thm2} that $\widehat{\Sigma}_M(d) =\Sigma_M(d)+o_p(1)$ and the condition that $\boldsymbol{\rho}$ is bounded, we can further replace $\widehat{\Sigma}_M^{-1}(\widehat{d})  $ in \eqref{thm3_p1_s2}, which yields that
  \begin{eqnarray}\label{thm3_p1_s3}  
 \eta_1=&& {\boldsymbol{\rho}}(\widehat{d})^\top  \widehat{\Sigma}_M^{-1}(\widehat{d})    \{\widehat{W}_{P}(\widehat{d})  - \widehat{W}_{U}(\widehat{d})  \}  + o_p(N_P^{-1/2}) \\\nonumber
 =&& {\boldsymbol{\rho}}(\widehat{d})^\top  \{ {\Sigma}_M^{-1}(\widehat{d}) +o_p(1)\} \{\widehat{W}_{P}(\widehat{d})  - \widehat{W}_{U}(\widehat{d})  \} + o_p(N_P^{-1/2}) \\\nonumber
  =&& {\boldsymbol{\rho}}(\widehat{d})^\top   {\Sigma}_M^{-1}(\widehat{d})   \{\widehat{W}_{P}(\widehat{d})  - \widehat{W}_{U}(\widehat{d})  \} +o_p(N_P^{-1/2}).
  \end{eqnarray} 
  
Combining \eqref{thm3_p1_s3}  with \eqref{pf_lem3_s1}, since $\boldsymbol{\rho}$ and ${\Sigma}_M^{-1}$ are bounded, with $0<t<+\infty$, we can show 
   \begin{eqnarray*}  
    \eta_1=&&{\boldsymbol{\rho}}(\widehat{d})^\top   {\Sigma}_M^{-1}(\widehat{d})   \{\widehat{W}_{P}(\widehat{d})  - \widehat{W}_{U}(\widehat{d})  \}+o_p(N_P^{-1/2}) \\\nonumber
   =&&  {\boldsymbol{\rho}}(\widehat{d})^\top   {\Sigma}_M^{-1}(\widehat{d}) \left\{ {1\over N_P}\sum_{i=1}^{N_P} \boldsymbol{w}^{(i)}_P(\widehat{d}) - {1\over N_U}\sum_{i=1}^{N_U} \boldsymbol{w}^{(i)}_U(\widehat{d}) + (1-\sqrt{t}) o_p(N_P^{-1/2})\right\}+o_p(N_P^{-1/2})\\\nonumber
   =&&  {\boldsymbol{\rho}}(\widehat{d})^\top   {\Sigma}_M^{-1}(\widehat{d}) \left\{ {1\over N_P}\sum_{i=1}^{N_P} \boldsymbol{w}^{(i)}_P(\widehat{d}) - {1\over N_U}\sum_{i=1}^{N_U} \boldsymbol{w}^{(i)}_U(\widehat{d})  \right\}+o_p(N_P^{-1/2}).
\end{eqnarray*}

Thus, \eqref{thm3_p1_eta1} is proved. This together with \eqref{thm3_p1_s1} yields \eqref{thm3_p1_mid1}. The proof of Part 1 is hence completed.

\bigskip
 
 \textbf{Proof of Part 2:} We next focus on  proving $
 {V}_n(\widehat{d}) =  {V}_n(d^{opt}) +o_p(N_P^{-1/2}).$ 
Define a class of function
\begin{eqnarray*}
&&\mathcal{F}_d(X_P,A_P,M_P,Y_P) = \Bigg[ {v}_P ( {d}) -  {\boldsymbol{\rho}}({d})^\top {\Sigma}_M^{-1}({d}) \left\{  \boldsymbol{w}_P({d})  - {1\over N_U}\sum_{j=1}^{N_U} \boldsymbol{w}^{(j)}_U({d}) \right\} \\
&&-{v}_P (d^{opt}) +  {\boldsymbol{\rho}}(d^{opt})^\top {\Sigma}_M^{-1}(d^{opt}) \left\{  \boldsymbol{w}_P(d^{opt})  - {1\over N_U}\sum_{j=1}^{N_U} \boldsymbol{w}^{(j)}_U(d^{opt}) \right\} : d(\cdot)\in \Pi \Bigg],
\end{eqnarray*}
where recall
\begin{eqnarray*}
{v}_P ( {d}) \equiv   {\mathbb{I}\{d(X_{P})\} [Y_{P} - {\mu}_P\{X_{P},d(X_{P})\} ] \over{A_{P}  {\pi}_P (X_{P})+(1-A_{P})\{1- {\pi}_P(X_{P})\}}}   +  {\mu}_P\{X_{P},d(X_{P})\}, 
\end{eqnarray*}
and
\begin{eqnarray*}
\boldsymbol{w}_P({d})  \equiv    {\mathbb{I}\{d(X_{P})\} [M_{P}  - \theta \{X_{P},d(X_{P})\} ]\over{A_{P} {\pi}_P (X_{P})+(1-A_{P})\{1-{\pi}_P(X_{P})\}}}   +  {\theta} \{X_{P},d(X_{P})\}  ,
\end{eqnarray*}
and $\Pi$ denotes the space of decision rules of interest such as $\Pi_1$ or  $\Pi_2$. 

Under the assumption (A5), we have the class of decision rules $\Pi$ is a Vapnik-Chervonenkis (VC) class of functions. By the conclusion of Lemma 2.6.18 in \cite{van1996weak}, we know the indicator function of a VC class of functions is still VC class. 

Furthermore, under assumptions (A5) and (A6), following results (iv) in Lemma A.1 of \cite{rai2018statistical}, it can be shown that ${v}_P ( {d}) $ and $\boldsymbol{w}_P({d})$ are continuous with respect to $d \in \Pi $. Therefore, we have $\mathcal{F}_d$ belongs to the VC class, and its entropy $\mathcal{J}(\mathcal{F}_d)$ is finite.

Define the supremum of the empirical process indexed by $ \mathcal{F}_d$ as
\begin{eqnarray*}
||\mathbb{G}_n||_{\mathcal{F}} \equiv &&\sup_{d \in \Pi} {1\over {\sqrt{N_P}}} \sum_{i=1}^{N_P}  \mathcal{F}_d(X_{P,i},A_{P,i},M_{P,i},Y_{P,i}) - \Mean\{ \mathcal{F}_d(X_{P},A_{P},M_{P},Y_{P}) \}\\
=&&\sup_{d \in \Pi} \sqrt{N_P} \{{V}_n( {d}) - {V}_n(d^{opt})-{V}({d}) +  {V}(d^{opt})\}.
\end{eqnarray*}

By the assumption (A6), we have  $\tilde{B}\equiv \max_{1\leq i\leq n}\mathcal{F}_d(X_{P,i},A_{P,i},M_{P,i},Y_{P,i})<\infty$. Define the asymptotic variance $\sigma_n^2 \equiv \sup_{d \in \Pi} P \mathcal{F}_d^2$, where $P$ is the common distribution of $\{X_P,A_P,M_P,Y_P\}$. Based on the central limit theorem, we have 
 \begin{equation*}  
\sqrt{N_P}\Big\{{V}_n(d)-V(d)\Big\}  \rightsquigarrow N\Big\{ 0, \sigma_0^2(d)  \Big\}, \text{ for all }{d \in \Pi} .
\end{equation*}
This implies  $\sigma_n^2 =O(N_P^{-1/2})$.

It follows from  the maximal inequality developed in Corollary 5.1 of \cite{chernozhukov2014gaussian} that there exist some constant $v_0\ge 1$ and $\tilde{C}>0$ such that
\begin{eqnarray*}
\Mean \Big[||\mathbb{G}_n||_{\mathcal{F}} \Big] \lesssim \sqrt{v_0 \sigma_n^2 \log\{\tilde{C}\mathcal{J}(\mathcal{F}_d)/\sigma_n\}} +{v_0 \tilde{B}^2\over\sqrt{N_P} }\log\{\tilde{C}\mathcal{J}(\mathcal{F}_d)/\sigma_n\}.
\end{eqnarray*}
The above right-hand-side is upper
 bounded by 
\begin{eqnarray*}
O(1)\sqrt{N_P^{-1/2}\log(N_P^{1/4})},
\end{eqnarray*}
 where $O(1)$ denotes some universal constant.

Hence, we have
\begin{eqnarray}\label{thm3_p2_mid1}
\sqrt{N_P} \{{V}_n( \widehat{d}) - {V}_n(d^{opt})-{V}(\widehat{d}) +  {V}(d^{opt})\} = o_p(1).
\end{eqnarray}

Under the margin condition (A8), following Theorem 2.3 in \citet{kitagawa2018should}, we have
\begin{eqnarray}\label{thm3_p2_mid2}
 {V}(\widehat{d}) =  {V}(d^{opt}) +o_p(N_P^{-1/2}).
\end{eqnarray}

Combining \eqref{thm3_p2_mid1} with \eqref{thm3_p2_mid2}, we have
\begin{eqnarray*}
 {V}_n(\widehat{d}) =  {V}_n(d^{opt}) +o_p(N_P^{-1/2}).
\end{eqnarray*}
This together with \eqref{thm3_p1_mid1} proves \eqref{thm3_p1_mid2}. Thus, we complete the proof of Part 2. 

\smallskip

 \textbf{Proof of Part 3:}
By the conclusion of Part 2, we have
 \begin{eqnarray*}  
\sqrt{N_P}\Big\{\widehat{V}(\widehat{d})-V(d^{opt})\Big\} =&& \sqrt{N_P}\Big\{\widehat{V}(\widehat{d})-{V}_n(d^{opt})\Big\} + \sqrt{N_P}\Big\{{V}_n(d^{opt})-V(d^{opt})\Big\}\\
 = &&o_p(1) + \sqrt{N_P}\Big\{{V}_n(d^{opt})-V(d^{opt})\Big\}.
\end{eqnarray*}

Thus, to prove the asymptotic normality of the proposed calibrated value estimator, it is sufficient to show the asymptotic normality of $\sqrt{N_P}\Big\{{V}_n(d^{opt})-V(d^{opt})\Big\}$ based on Slutsky's theorem. 

Noticing
 \begin{eqnarray*}  
{V}_n(d^{opt})=&&{1\over N_P}\sum_{i=1}^{N_P} {v}_P^{(i)}(d^{opt}) -  {\boldsymbol{\rho}}(d^{opt})^\top {\Sigma}_M^{-1}(d^{opt}) \left\{ {1\over N_P}\sum_{i=1}^{N_P} \boldsymbol{w}^{(i)}_P(d^{opt})  - {1\over N_U}\sum_{j=1}^{N_U} \boldsymbol{w}^{(j)}_U(d^{opt}) \right\},\\
=&&{1\over N_P}\sum_{i=1}^{N_P} \left[{v}_P^{(i)}(d^{opt}) -  {\boldsymbol{\rho}}(d^{opt})^\top {\Sigma}_M^{-1}(d^{opt}) \left\{   \boldsymbol{w}^{(i)}_P(d^{opt})  - {1\over N_U}\sum_{j=1}^{N_U} \boldsymbol{w}^{(j)}_U(d^{opt}) \right\} \right],
\end{eqnarray*}
where ${\boldsymbol{\rho}}(d^{opt})^\top {\Sigma}_M^{-1}(d^{opt}) $ is a fixed constant.

By the central limit theorem with $T=\text{lim}_{N_P\to +\infty} t < +\infty$, we have 
 \begin{equation*}  
\sqrt{N_P}\Big\{{V}_n(d^{opt})-V(d^{opt})\Big\}  \rightsquigarrow N\Big\{ 0, \sigma^2(d^{opt}) \Big\}. 
\end{equation*}

Next, we give the explicit form of $\sigma^2(d^{opt})$. Notice that \begin{eqnarray*}  
\sqrt{N_P}\Big\{{V}_n(d^{opt})-V(d^{opt})\Big\}  =&&\underbrace{{1\over \sqrt{N_P}}\sum_{i=1}^{N_P} \left\{{v}_P^{(i)}(d^{opt}) -V(d^{opt})\right\}}_{\eta_2}\\
&&-\underbrace{{1\over \sqrt{N_P}}\sum_{i=1}^{N_P}   {\boldsymbol{\rho}}(d^{opt})^\top {\Sigma}_M^{-1}(d^{opt}) \left\{   \boldsymbol{w}^{(i)}_P(d^{opt}) - W(d^{opt}) \right\}}_{\eta_3}\\
&&  +\underbrace{ {\sqrt{t}\over \sqrt{N_U}}\sum_{j=1}^{N_U}  {\boldsymbol{\rho}}(d^{opt})^\top {\Sigma}_M^{-1}(d^{opt}) \left\{  \boldsymbol{w}^{(j)}_U(d^{opt})-W(d^{opt}) \right\}}_{\eta_4}.
\end{eqnarray*}

Since the two samples ($P$ and $U$) are independently collected from two different separate studies, we have the independent and identically distributed double robust terms $\{{v}_P^{(i)}(d^{opt}) ,\boldsymbol{w}^{(i)}_P(d^{opt}) \}_{1\leq i \leq N_P}$ are independent of $\{\boldsymbol{w}^{(i)}_U(d^{opt}) \}_{1\leq i \leq N_U}$. Hence, the variance of $\sqrt{N_P}\Big\{{V}_n(d^{opt})-V(d^{opt})\Big\} $ is 
\begin{eqnarray}\label{thm3_p3_mid1}  
\sigma^2 =\lim_{N_P\to \infty} \left\{\Var(\eta_2)+ \Var(\eta_3) + \Var(\eta_4)-2 \Cov(\eta_2,\eta_3) \right\} .
\end{eqnarray}

Using similar arguments in proving results (ii) and (iii) in Theorem \ref{thm2}, we have
\begin{eqnarray}\label{thm3_p3_mid2}  
	\left\{\begin{array}{ll}
		\lim_{N_P\to \infty}\Var(\eta_2) =  \sigma_Y^2(d^{opt})
		\\
		\lim_{N_P\to \infty}\Var(\eta_3) = \boldsymbol{\rho}(d^{opt})^\top\Sigma_M^{-1} (d^{opt}) \Sigma_P(d^{opt}) \Sigma_M^{-1} (d^{opt})\boldsymbol{\rho}(d^{opt})\\
		\lim_{N_P\to \infty}\Var(\eta_4) = \boldsymbol{\rho}(d^{opt})^\top\Sigma_M^{-1} (d^{opt})T\Sigma_U(d^{opt})\Sigma_M^{-1} (d^{opt})\boldsymbol{\rho}(d^{opt})\\
		\lim_{N_P\to \infty} \Cov(\eta_2,\eta_3) =  \boldsymbol{\rho}(d^{opt})^\top\Sigma_M^{-1}(d^{opt}) \boldsymbol{\rho}(d^{opt}).\\
	\end{array}
	\right.
\end{eqnarray} 

Combining \eqref{thm3_p3_mid1} with \eqref{thm3_p3_mid2}, using the definition that $\Sigma_M(d) = \Sigma_P(d)+ T\Sigma_U(d)$, we have
\begin{eqnarray*}
\sigma^2 = \sigma_Y^2(d^{opt}) - \boldsymbol{\rho}(d^{opt})^\top\Sigma_M^{-1}(d^{opt}) \boldsymbol{\rho}(d^{opt})  . 
\end{eqnarray*}

Thus, we have 
 \begin{equation*}  
\sqrt{N_P}\Big\{{V}_n(d^{opt})-V(d^{opt})\Big\}  \rightsquigarrow N\Big[ 0, \sigma_Y^2(d^{opt}) - \boldsymbol{\rho}(d^{opt})^\top\Sigma_M^{-1}(d^{opt}) \boldsymbol{\rho}(d^{opt})  \Big]. 
\end{equation*}
The proof is hence completed.

\subsection{Proof of Theorem \ref{thm3_hete}}

The proof of Theorem \ref{thm3_hete} follows the proof of Theorem \ref{thm3}. With a similar manner, we can show the asymptotic normality of $\sqrt{N_P}\{\widehat{V}_{R}(\widehat{d}_R)-V(d^{opt}) \}$ by three parts specified in Section \ref{sec:pf_thm3}. The only difference is to replace the previous value difference between two samples ($\{\widehat{W}_{P}(d) - \widehat{W}_{U}(d)\} $) with the rebalanced value difference based on joint sample ($\{\widehat{W}_{1}(d) - \widehat{W}_{0}(d)\} $) with a sample ratio, under the consistency results (vi), (v), and (vii) in Theorem \ref{thm2}. We omit the details for brevity.

\end{document}